\documentclass[%
 reprint,
nofootinbib,
 amsmath,amssymb,
 aps,
twocolumn,
floatfix
]{revtex4-2}

\usepackage{graphicx,smartdiagram}
\usepackage{dcolumn}
\usepackage{bm}
\usepackage{amsmath,amssymb,nccmath}
\usepackage{xcolor}
\usepackage{ytableau}
\usepackage[colorlinks,linkcolor=blue]{hyperref}
\usepackage{tikz}
\usepackage{orcidlink}
\usepackage{academicons}
\definecolor{orcidlogocol}{HTML}{A6CE39}
\newcommand{\orcid}[1]{\href{https://orcid.org/#1}{\textcolor[HTML]{A6CE39}{\aiOrcid}}}

\begin{document}

\preprint{XYZ}

\title{Spacetime-symmetry breaking effects in gravitational-wave generation at the first post-Newtonian order}

\author{Nils A. Nilsson\,\orcidlink{0000-0001-6949-3956}}
\email{nils.nilsson@obspm.fr}

\author{Christophe Le-Poncin Lafitte\,\orcidlink{0000-0002-3811-1828}}%
 \email{christophe.leponcin@obspm.fr}
\affiliation{SYRTE, Observatoire de Paris, Universit\'e PSL, CNRS, Sorbonne Universit\'e,
LNE, 61 avenue de l'Observatoire, 75014 Paris, France}%

\date{\today}

\begin{abstract} 
Current searches for signals of departures from the fundamental symmetries of General Relativity using gravitational waves are largely dominated by propagation effects like dispersion and birefringence from highly dynamic sources such as coalescing binary-black holes and neutron stars. In this paper we take steps towards probing the nature of spacetime symmetries in the {\it generation-stage} of gravitational waves; by using a generic effective-field theory, we solve the modified Einstein equations order-by-order (in the coefficients for the symmetry breaking) for a generic source, and we write down the the first Post-Newtonian corrections, which includes contributions from the spacetime-symmetry breaking terms. Choosing as the source a system of point particles allows us to write down a simple toy solution explicitly, and we see that in contrast to General Relativity, the monopolar and dipolar contributions are non-vanishing. We comment on the detectability of such signals by the Laser Interferometer Space Antenna (LISA) space mission, which has high signal-to-noise galactic binaries (which can be modelled as point particles) well inside its predicted sensitivity band, sources which are inaccessible for current ground-based detectors, and we also discuss the possibility of going beyond the quadrupole formula and the first Post-Newtonian order, which would reveal effects which could be probed by ground-based detectors observing coalescence events.
\end{abstract}

\maketitle

\tableofcontents

\section{Introduction}
The detection of gravitational waves by the Laser Interferometer Gravitational-Wave Observatory (LIGO) and Virgo collaborations not only confirmed long sought-after predictions by General Relativity (GR), but also opened an entirely new window to the Universe through gravitational-wave astronomy. With the detections of binary coalescence events soon numbering $10^2$, a number which is expected to skyrocket during the fourth Observing Run (O4)\footnote{Which, at the time of writing this paper, has recently begun.}, gravitational waves are now able to probe the nature of the gravitational interaction in truly extreme environments; moreover, since they travel virtually unimpeded through space, high-precision bounds on the propagation speed of gravity as compared to photons are now possible \cite{LIGOScientific:2017zic}. On the astrophysical side, gravitational waves have enabled discoveries such as the existence of heavy stellar-mass black holes and precise constraints on the mass and radius of neutron stars \cite{Yagi:2015pkc,Chatziioannou:2018vzf,Radice:2017lry}. 

When searching for a theory unifying GR and the Standard Model of particle physics (SM), it may be necessary to relax some of the underlying tenets linking the two paradigms; indeed, many proposals of quantum gravity predict (or allow) that Lorentz symmetry and CPT (Charge-Parity-Time) symmetry are not exact at energy scales relevant for quantum gravity \cite{Kostelecky:1988zi,Kostelecky:1991ak,Gambini:1998it,Carroll:2001ws,Addazi:2021xuf,Mariz:2022oib,Horava:2009uw}, and signals of such {\it spacetime-symmetry breaking} may be accessible to highly sensitive tests. Since Lorentz symmetry is a combination of rotation and boost invariance, its breaking results in non-standard preferred directions and velocity-dependent quantities. Using gravitational-wave observations, several tests of GR have been performed, for example \cite{LIGOScientific:2017zic,LIGOScientific:2021sio,Liu:2020slm}, which so far has revealed no departure from known physics. Given that GR holds to very high accuracy, any spacetime-symmetry breaking in nature must be very small at the energy scales available to us, and with very little experimental guidance to direct theoretical model building, a practical approach is to search for features of the underlying theory through effective-field theory. In the past decades, an effective-field theory known as the Standard-Model Extension (SME) has emerged as one of the de-facto standard tools for testing Lorentz and CPT symmetry; the SME contains GR, SM, and every possible local Lorentz, CPT, and/or diffeomorphism-breaking term suppressed by increasing inverse orders of the Planck mass \cite{Colladay:1996iz,Colladay:1998fq,Kostelecky:2003fs,Bailey:2006fd,Kostelecky:2010ze,Bluhm:2014oua,Kostelecky:2017zob,Kostelecky:2020hbb,ONeal-Ault:2020ebv}. The SME has inspired a large research effort, with constraints on the gravitational sector being obtained using gravitational waves \cite{Haegel:2022ymk,Kostelecky:2016kfm,Kostelecky:2015dpa,LIGOScientific:2017zic,Wang:2021ctl,Niu:2022yhr,Nilsson:2022mzq}, Solar-system tests \cite{Iorio:2012gr,Hees:2015mga,LePoncin-Lafitte:2016ocy}, lunar laser ranging \cite{Bourgoin:2016ynf,Bourgoin:2017fpo,Bourgoin:2020ckq}, pulsar tests \cite{Shao:2014oha,Shao:2018vul}, and many more. An exhaustive list\footnote{Updated annually.} of all constraints obtained on SME coefficients in all sectors can be found in \cite{Kostelecky:2008ts}. A generic term in the SME gravitational sector can be written schematically as
$$
    \mathcal{L}_{\rm SME} \supseteq \text{ (SME coefficient) } \times \text{ (dimension-$d$ operator)},
$$
and is built by contracting a conventional mass dimension-$d$ operator with a coefficient tensor which parameterises Lorentz and/or CPT violation; these coefficients transform as scalars under so-called {\it particle rotations} \cite{Kostelecky:2003fs}, and to lowest mass dimension in the gravitational sector (mass dimension $d=4$), the relevant curvature quantity is the Riemann tensor. The values of the SME coefficients are then put to experimental tests, where any non-zero value indicates that the spacetime symmetries are broken. Since known physics is contained within the SME, and with the symmetry-breaking terms being constructed from quite generic operators, specific models can be realised by specifying the values of the SME coefficients, and several maps to models have been found, for example~\cite{Bailey:2018ifc} (non-commutative gravity) and \cite{ONeal-Ault:2020ebv} (Ho\v{r}ava-Lifshitz gravity).  

In general, there are two mechanisms through which non-zero SME coefficients can arise, called {\it spontaneous} and {\it explicit} breaking \cite{Bluhm:2014oua}: in the case of spontaneous breaking, the underlying Lagrange density is still Lorentz invariant and the symmetry breaking occurs through a Higgs-like mechanism at the level of the Euler-Lagrange equations\footnote{With the associated Nambu-Goldstone modes and massive modes~\cite{Bluhm:2010mi}.}. In this case, the SME coefficients need to be considered as dynamical fields with equations of motion of their own. In contrast, should the symmetry breaking arise explicitly, the coefficients are instead ``fixed'' at the level of the action and play no dynamical role\footnote{Where it should be noted that the Lagrangian itself is no longer Lorentz invariant.}. In general, explicit symmetry breaking in the gravitational sector is {\it incompatible} with Riemannian geometry through highly non-trivial complications at the level of the Bianchi identities~\cite{Kostelecky:2003fs}, but exceptions exist, see for example \cite{ONeal-Ault:2020ebv,Nilsson:2022mzq,Khodadi:2023ezj,Reyes:2022mvm}. In this work, we shall focus our attention on the case of spontaneous origin of the symmetry breaking, which is compatible with several theoretical scenarios, for example string theory \cite{Kostelecky:1988zi, Kostelecky:1991ak}, random dynamics \cite{Froggatt:2002gf}, and more (see for example \cite{Gambini:1998it,Mavromatos:2004sz,Amelino-Camelia:2002aqz}). A well-known example of a candidate quantum-gravity model which incorporates explicit spacetime-symmetry breaking is Ho\v{r}ava-Lifshitz gravity \cite{Horava:2009uw}. 

The Laser Interferometer Space Antenna (LISA) mission\footnote{\url{https://www.elisascience.org/}} is the European Space Agency future space based gravitational-wave detector, which will be highly sensitive to low-frequency gravitational waves in the band $< 10^{-4}$ Hz to $>10^{-1}$ Hz \cite{Barausse:2020rsu,LISA:2022kgy}. Within this band lie a multitude of Galactic sources comprised of white dwarfs and neutron stars in different combinations, known as Galactic Binaries. These non-coalescing, relatively slow-moving sources emit continuous, quasi-monochromatic gravitational waves with a period of minutes to hours which will be observable by LISA throughout the entire mission lifetime \cite{Cornish:2017vip}. The fact that these are ``weak'' and slow-moving sources means that they can be treated using a Post-Newtonian expansion, without the need to employ numerical relativity and computationally expensive waveform modelling. A number of galactic binaries with exceptional signal-to-noise ratio are known as verification binaries, which are guaranteed sources for LISA, with simulations predicting $\mathcal{O}(10^4)$ sources within the Milky Way \cite{Cornish:2017vip}; the use of sources within our own galaxy also precludes the need to take cosmology into account. These sources are of significantly lower energy than the mergers detected by ground-based detectors, but they are plentiful and continuously observable, and so the amount of statistics which LISA can gather will be considerable. The formalism developed in this paper would be well suited for tests with such sources (see Section \ref{sec:sol}), but can also be used for the inspiral phase of binary coalescence events as observed by ground-based detectors.

A common denominator for most spacetime-symmetry tests with gravitational waves is that they rely on {\it propagation effects} such as birefringence and dispersion through the use of modified dispersion relations of the form
$$
\omega = |\mathbf{p}|(1\pm\rm{corrections}),
$$
in terms of the components of the 4-momentum $p^\mu=(\omega,\mathbf{p})$; in this paper, we outline the steps for obtaining constraints from the {\it generation stage} of gravitational waves as a complement to the more common propagation studies\footnote{It should be noted that modified generation has been partially explored in a vector subset of the SME known as the bumblebee model \cite{Amarilo:2019lfq,Amarilo:2023wpn}.}. Starting with a generic source, we write down the field equations in the presence of spacetime-symmetry breaking terms at arbitrary order in mass dimension $d$. We later specify the source as a system of point particles and provide sample solutions for a simple case, and we briefly comment on their detectability with the future LISA space mission, as well as 
 with current ground-based detectors. 
 
 The paper is organised as follows: In Section~\ref{sec:theory}, we write down the theoretical framework and formal solutions to the relaxed field equations; in Section~\ref{sec:PN}, we perform the Post-Newtonian (PN) expansion in the near zone and discuss the PN order required in our solutions; in Section~\ref{sec:grsolint} we discuss the contributions from the wavezone integrals; in Section~\ref{sec:quadrupole} we introduce the quadrupole formula for a generic source; in Section~\ref{sec:sol} we present a simple sample solution to our equations for point particles; we briefly discuss the possibilities of detection by LISA and ground-based telescopes, outline future work, and conclude in Section~\ref{sec:disc}. We use Greek letters ($\mu,\nu,\hdots$) for spacetime indices and Latin letters ($i,j,k,\hdots$) to denote spatial indices. When referring to mass dimension $d$ we use natural units where $c=\hbar=1$, but we write out $c$ explicitly in our equations, as it helps the counting of Post-Newtonian orders. In this paper, we employ several notions of perturbative orders; we have made every attempt at clarity, but caution is needed. In places where we do not specify the type of order, it should be understood that we are referring to the order-by-order solution scheme which we adopt in Eq.~(\ref{eq:orderbyorder}).

\section{Theoretical framework}\label{sec:theory}
We define the metric potentials as fluctuations around the Minkowski metric\footnote{We use the east-coast signature $(-+++)$.} as
\begin{equation}
    h_{\alpha\beta} \equiv \eta_{\alpha\beta} - g_{\alpha\beta}.
\end{equation}

Ensuring that the usual linear gauge symmetry $h_{\alpha\beta}\to h_{\alpha\beta}+\partial_\alpha\xi_\beta+\partial_\beta\xi_\alpha$ is satisfied, we can write the general and complete Lagrange density quadratic in the metric potentials $h_{\alpha\beta}$ as \cite{Kostelecky:2016kfm}
\begin{equation}
\begin{aligned}
    \mathcal{L} =& \tfrac{1}{8\kappa}\epsilon^{\mu\rho\alpha\kappa}\epsilon^{\nu\sigma\beta\lambda}\eta_{\kappa\lambda}h_{\mu\nu}\partial_\alpha\partial_\beta h_{\rho\sigma} \\&+ \tfrac{1}{8\kappa}h_{\mu\nu}\left(\hat{s}^{\mu\rho\nu\sigma}+\hat{q}^{\mu\rho\nu\sigma}+\hat{k}^{\mu\rho\nu\sigma}\right)h_{\rho\sigma},
    \end{aligned}
\end{equation}
which includes GR and all spacetime-symmetry breaking terms at arbitrary mass-dimension $d$. Here,  $\kappa=8\pi G$, $\epsilon^{\mu\rho\alpha\kappa}$ is the totally antisymmetric Levi-Civita tensor density, and the first term gives the linearised Einstein tensor for GR (to first order in $h$). The hatted quantities are the gauge-invariant SME operators defined as
\begin{equation}
\begin{aligned}
    \hat{s}^{\mu\rho\nu\sigma} &=& s^{(d)~\mu\rho\epsilon_1\nu\sigma\epsilon_2\hdots\epsilon_{d-2}}\partial_{\epsilon_1}\hdots\partial_{\epsilon_{d-2}},\\
    \hat{q}^{\mu\rho\nu\sigma} &=& q^{(d)~\mu\rho\epsilon_1\nu\epsilon_2\sigma\epsilon_3\hdots\epsilon_{d-2}}\partial_{\epsilon_1}\hdots\partial_{\epsilon_{d-2}},\\
    \hat{k}^{\mu\rho\nu\sigma} &=& k^{(d)~\mu\epsilon_1\nu\epsilon_2\rho\epsilon_3\sigma\epsilon_4\hdots\epsilon_{d-2}}\partial_{\epsilon_1}\hdots\partial_{\epsilon_{d-2}},
    \end{aligned}
\end{equation}
where $\hat{s}^{\mu\rho\nu\sigma}$ is CPT even with $d\geq4$, $\hat{q}^{\mu\rho\nu\sigma}$ is CPT odd with $d\geq5$, and $\hat{k}^{\mu\rho\nu\sigma}$ is CPT even with $d\geq6$. The symmetry properties of the hatted operators can be read off the Young tableaux in Figure~\ref{fig:yt}.

\ytableausetup{centertableaux}
\begin{figure}[ht!]
\begin{ytableau}
\mu & \nu & \hdots \\
\rho & \sigma \\
\circ & \circ
\end{ytableau}
\qquad
\begin{ytableau}
    \mu & \nu & \sigma & \hdots\\
    \rho & \circ & \circ \\
    \circ
\end{ytableau}
\qquad
\begin{ytableau}
    \mu & \nu & \rho & \sigma & \hdots \\
    \circ & \circ & \circ & \circ
\end{ytableau}
\caption{Young tableaux for $s^{(d)\mu\rho\circ\nu\sigma{\circ\circ^{d-4}}}$ (left), $q^{(d)\mu\rho\circ\nu\circ\sigma{\circ\circ^{d-5}}}$ (middle), and $k^{(d)\mu\circ\nu\circ\rho\circ\sigma{\circ\circ^{d-6}}}$ (right).}
\label{fig:yt}
\end{figure}
Since they are in essence duals (of codimension-2) to the widely used barred SME coefficients of linearised gravity, it is possible to map between them using \cite{Kostelecky:2016kfm}
\begin{equation}\label{eq:hatbar}
    \hat{s}^{\mu\rho\nu\sigma} = -\epsilon^{\mu\rho\alpha\kappa}\epsilon^{\nu\sigma\beta\lambda}\bar{s}_{\kappa\lambda}\partial_\alpha\partial_\beta,
\end{equation}
and similarly for $\hat{q}^{\mu\rho\nu\sigma}$ and $\hat{k}^{\mu\rho\nu\sigma}$. Therefore, a single component of a hatted operator represents a specific {\it combination} of barred coefficients with associated partial derivatives, which together make up irreducible pieces of the hatted operators. 

The Euler-Lagrange equations read
\begin{equation}\label{eq:eeq}
    G_L^{\mu\nu}+M^{\mu\nu\rho\sigma}h_{\rho\sigma} - \frac{\kappa}{c^4} \tau^{\mu\nu} = 0,
\end{equation}
where $\tau^{\mu\nu}$ is the matter stress-energy tensor and $G_L^{\mu\nu}$ is the linearised Einstein tensor
\begin{equation}
    G_L^{\mu\nu} = -\tfrac{1}{2}\eta_{\rho\sigma}\epsilon^{\mu\rho\alpha\kappa}\epsilon^{\nu\sigma\beta\lambda}\partial_\alpha\partial_\beta h_{\kappa\lambda},
\end{equation}
and $M^{\mu\nu\rho\sigma}$ is
\begin{equation}\label{eq:Mdef}
\begin{aligned}
    M^{\mu\nu\rho\sigma} =& -\tfrac{1}{2}\Big[\tfrac{1}{2}\left(\hat{s}^{\mu\rho\nu\sigma}+\hat{s}^{\mu\sigma\nu\rho}\right)+\hat{k}^{\mu\rho\nu\sigma}\\&+\tfrac{1}{4}\left(\hat{q}^{\mu\rho\nu\sigma}+\hat{q}^{\nu\rho\mu\sigma}+\hat{q}^{\mu\sigma\nu\rho}+\hat{q}^{\nu\sigma\mu\rho}\right)\Big],
    \end{aligned}
\end{equation}
which is symmetric in the first and last pairs of indices. Similar modifications to the Einstein equations were found in the context of Chern-Simons gravity, where the non-standard terms were recast as a modified dynamical matter source~\cite{Alexander:2009tp}.

In order to make the field equations more tractable, we introduce the trace-reversed metric potentials (henceforth denoted with a bar) as
\begin{equation}
    h_{\alpha\beta}=\bar{h}_{\alpha\beta}-\tfrac{1}{2}\bar{h}\eta_{\alpha\beta}, \quad h = - \bar{h},
\end{equation}
and we note that the field equations expressed using the trace-reversed potentials are equivalent to the first-order (in $\bar{h}$) limit of the relaxed Einstein equations in the Landau-Lifshitz formulation of general relativity, i.e. using the gothic metric $\mathfrak{g}^{\mu\nu} = \sqrt{-g}g^{\mu\nu}$ and
\begin{equation}
    \bar{h}^{\alpha\beta} \equiv \eta^{\alpha\beta}-\mathfrak{g}^{\alpha\beta} + \mathcal{O}(h^2).
\end{equation}
Thanks to this equivalence, we will be able to use the powerful machinery presented in the book \cite{poisson_will_2014} in the following sections, the limitation being that we may only consider terms which are first order in $\bar{h}^{\mu\nu}$. In the Einstein equations, this appends two terms to the energy momentum tensor (now a pseudotensor), which now reads
\begin{equation}
    \tau^{\mu\nu} = T^{\mu\nu}+\tau_{\rm H}^{\mu\nu} + \tau_{\rm LL}^{\mu\nu},
\end{equation}
where $T^{\mu\nu}$ is the stress-energy tensor of the source, $\tau_{\rm H}^{\mu\nu}$ and $\tau_{\rm LL}^{\mu\nu}$ is the harmonic gauge and Landau-Lifshitz contribution to the energy-momentum pseudotensor, respectively. In GR, $\tau^{\mu\nu}_{\rm H}$ and $\tau^{\mu\nu}_{\rm LL}$ contain terms at second order in the metric potential $h$ such as $\partial h \partial h$ and $h\partial h$, but since our approach is only valid to linear order in $h$, we must discard these quadratic contributions.
In the vacuum case, plane-wave solutions to the modified Einstein equations (\ref{eq:eeq}) in momentum space admit a dispersion relation of the form $\omega = |\mathbf{p}|(1-\zeta^0\pm\zeta^1)$, where $\zeta^{0, 1}$ consist of contracted combinations of $\hat{s}^{\mu\rho\nu\sigma}$, $\hat{q}^{\mu\rho\nu\sigma}$, and $\hat{k}^{\mu\rho\nu\sigma}$ suppressed by higher inverse orders of momentum $|\mathbf{p}|$ \cite{Mewes:2019dhj,Haegel:2022ymk}. Here, the $\pm$ sign shows the appearance of birefringence of the propagating modes, an effect which occurs for odd mass-dimension $d\geq 5$ operators. Due to the highly suppressed nature of the propagation effects, cosmological distances are normally required for the effects to build up sufficiently, as was used for gravitational waves in \cite{Haegel:2022ymk,ONeal-Ault:2021uwu,Mewes:2019dhj} and for photon propagation in \cite{Amelino-Camelia:1997ieq, Kifune:1999ex}. In this paper, we focus on generation effects only, although corrections from propagation can in principle be applied to the resulting waveforms when considering extragalactic sources.

Considering small departures from the symmetries of GR to linear order in the metric potentials, we now solve Eq.~(\ref{eq:eeq}) expressed in terms of the trace-reversed potentials $\bar{h}^{\mu\nu}$ by splitting the potential $\bar{h}^{\mu\nu}$ into two parts as\footnote{This should not be confused with the Post-Minkowskian expansion which generally uses similar notation.}
\begin{equation}\label{eq:orderbyorder}
    \bar{h}^{\mu\nu} = \bar{h}^{(0)\mu\nu} + \bar{h}^{(1)\mu\nu},
\end{equation}
where $\bar{h}^{(0)\mu\nu}$ is the trace-reversed GR solution and $\bar{h}^{(1)\mu\nu}$ contains the symmetry-breaking terms, which is a similar approach to that of \cite{Kostelecky:2016kfm}. We choose the harmonic gauge, which with the above equation in mind reads
\begin{equation}\label{eq:gauge}
    \partial_\mu \bar{h}^{\mu\nu}=\partial_\mu\left( \bar{h}^{(0)\mu\nu}+\bar{h}^{(1)\mu\nu}\right) = 0.
\end{equation}
Since the GR metric potential satisfies this condition on its own, $\partial_\mu \bar{h}^{(0)\mu\nu} = 0$, the above choice implies that the harmonic gauge condition also holds at first order $\partial_\mu h^{(1)\mu\nu}=0$.

The equations of motion for the potentials read
\begin{equation}\label{eq:waveqs}
    \begin{aligned}
        \Box \bar{h}^{(0)\mu\nu} =& -\frac{2\kappa}{c^4}\tau^{\mu\nu} \\
        \Box \bar{h}^{(1)\mu\nu} =& 2\bar{M}^{\mu\nu\rho\sigma}\bar{h}^{(0)}_{\rho\sigma},
    \end{aligned}
\end{equation}
with the GR solution acting as the source for first order $\bar{h}^{(1)\mu\nu}$;
the full solution will be the sum of the two contributions. Here, we have introduced the trace-reversed $\bar{M}^{\mu\nu\rho\sigma}$ and the trace-reversal operator $\mathcal{A}_{\kappa\lambda}^{~~\rho\sigma}$ following \cite{Kostelecky:2016kfm} as
\begin{equation}
    \begin{aligned}
        \bar{M}^{\mu\nu\rho\sigma} &= M^{\mu\nu\kappa\lambda} \mathcal{A}_{\kappa\lambda}^{~~\rho\sigma}, \\
        \mathcal{A}_{\kappa\lambda}^{~~\rho\sigma} &= \tfrac{1}{2}\left(\eta_\kappa^{~\rho}\eta_\lambda^{~ \sigma}+\eta_\kappa^{~\sigma}\eta_\lambda^{~\rho}-\eta_{\kappa\lambda}\eta^{\rho\sigma}\right).
    \end{aligned}
\end{equation}

At the GR level, the formal solution reads
\begin{equation}\label{eq:GRh}
    \bar{h}^{(0)\mu\nu}(x) = \frac{\kappa}{4\pi c^4}\int d^4y \, G(x-y) \tau^{\mu\nu}(y),
\end{equation}
where $G(x-y)$ is the retarded Green's function associated with the Minkowski d'Alembertian operator $\Box \equiv \partial^\alpha\partial_\alpha$ defined as
\begin{equation}\label{eq:green}
    \Box G(x-y) = -4\pi\delta^{(4)}(x-y).
\end{equation}
By inserting the GR solution (\ref{eq:GRh}) into the first-order wave equation (\ref{eq:waveqs}), the full solution can now be schematically written as
\begin{equation}\label{eq:fullsol}
    \boxed{\bar{h}^{(1)\mu\nu} = -\frac{\kappa}{8\pi^2c^4}\int d^4yd^4zG(x-y)G(y-z)\bar{M}^{\mu\nu\alpha\beta}\tau_{\alpha\beta}(z)},
\end{equation}
which bears some similarity to integrals appearing in loop calculations~\cite{peskin1995introduction} and tails-of-memory effects~\cite{Trestini:2023wwg}. Eq.~(\ref{eq:fullsol}) is a troublesome integral with potentially non-local and acausal pieces; this can be seen by applying the d'Alembertian to the second equation in (\ref{eq:waveqs}), leading to $\Box^2\bar{h}^{(1)}\sim \bar{M}\tau$. The right-hand side of this equation is a source with compact support, but the left-hand side is a non-local operator, which was discussed in some detail in \cite{Pais:1950za, Bailey2023}. At zeroth order (the GR solution), the retarded solution must be considered the physical one, and we apply the same logic at first order, and only focus on the retarded solutions.
Moreover, since the source in the first-order wave equation in (\ref{eq:waveqs}) is the GR wave solution $h^{(0)}$, this equation does not have a compact source, as $h^{(0)}$ is defined over all space.
The rest of the paper will be devoted to finding ways to evaluate the integral in Eq.~(\ref{eq:fullsol}).

\subsection{GR solution}
The solution to Eq.~(\ref{eq:green}) gives the retarded (and advanced) Green's function related to the inverse Minkowski d'Alembertian, which we can write as
\begin{equation}
    G(x-x^\prime) \equiv \frac{\delta^{(3)}{((ct-ct^\prime)-|\mathbf{x}-\mathbf{x}^\prime|)}}{|\mathbf{x}-\mathbf{x}^\prime|},
\end{equation}
which simplifies Eq.~(\ref{eq:GRh}) to
\begin{equation}\label{eq:zeroth}
    \bar{h}^{(0)\mu\nu}(x) = \frac{\kappa}{4\pi c^4}\int d^3x^\prime \frac{\tau^{\mu\nu}(\tau,\mathbf{x^\prime})}{|\mathbf{x}-\mathbf{x}^\prime|},
\end{equation}
where $\tau$ denotes the retarded time coordinate, defined as 
\begin{equation}
\tau  \equiv t- \frac{1}{c}|\mathbf{x}-\mathbf{x^\prime}|,
\end{equation}
where $|\mathbf{x}-\mathbf{x^\prime}|$ is the Euclidean distance. We integrate Eq.~(\ref{eq:zeroth}) over the entire past lightcone of the point $x$, which we call $\mathcal{C}(x)$. Now, it is possible to split the integral over the {\it near zone} $\mathcal{N}(x)$ and the {\it wave zone} $\mathcal{W}(x)$ (see Figure~\ref{fig:lightcone} for definitions of the integration regions) of the source as
\begin{equation}
    \begin{aligned}
        \bar{h}^{(0)\mu\nu}(x) =& \frac{\kappa}{4\pi c^4}\int_{\mathcal{N}(x)} d^3x^\prime \frac{\tau^{\mu\nu}(\tau,\mathbf{x^\prime})}{|\mathbf{x}-\mathbf{x}^\prime|}\\&+\frac{\kappa}{4\pi c^4}\int_{\mathcal{W}(x)} d^3x^\prime \frac{\tau^{\mu\nu}(\tau,\mathbf{x^\prime})}{|\mathbf{x}-\mathbf{x}^\prime|}, 
    \end{aligned}
\end{equation}
where the wave-zone integral can safely be neglected to the level of accuracy we require in this paper; the remaining integral is over the near zone $\mathcal{N}(x)$. We stress here that we will place the field point in $\mathcal{W}(x)$ for the first-order solution $\bar{h}^{(1)}$, but since this will involve an integral of $\bar{h}^{(0)}$ over $\mathcal{N}(x)$, the ``source'' (which contains the GR solution) must be evaluated with the field point in the near zone. Once we have the first-order solution for a wave-zone field point, we add it to the known GR solution evaluated for the same field point.

Since we will be manipulating the source and field points, we introduce the following notation to denote their locations to avoid confusion: for example, $\int_{\mathcal{N}_\mathcal{W}(x)}$
is an integral over the {\it near zone} when the field point $x$ is in the wave zone, and so on. A diagram showing the different integration regions can be seen in Figure~\ref{fig:lightcone}.
In the following sections, we will solve the above integrals using a Post-Newtonian (PN) expansion. 
\def\b{2}
\def\h{2}
\begin{center}
    \begin{figure}[h]
    \begin{tikzpicture}[scale=2]
\foreach \p [count=\i from 0,
    evaluate={\rx=\b; \ry=\rx*\p/2; \ta=90-atan2(\h,\ry);}]
  in {0.3}{
\begin{scope}[shift={({mod(\i,3)*\b*1.25},{-floor(\i/3)*\h*1.25})}]
\fill [gray!20]
(0, \h) -- (\ta:\rx+0 and \ry) arc (\ta:180-\ta:\rx+0 and \ry) -- cycle;
\fill [gray!30] ellipse [x radius=\rx, y radius=\ry];
\draw [dashed] (\ta:\rx+0 and \ry) arc (\ta:180-\ta:\rx+0 and \ry);
\draw (0, \h) -- (\ta:\rx+0 and \ry) arc (\ta:-180-\ta:\rx+0 and \ry) -- cycle;

\draw (-0.75,0) ellipse (0.5 and 0.2);
\draw [dashed] (-1.25,0) -- (-1.25,0.77);
\draw (-1.25,0.78) -- (-1.25,2);
\draw (-1.25,2) arc (180:360:0.5 and 0.2);
\draw [dashed] (-0.25,0) -- (-0.25,1.76);
\draw (-0.25,1.77) -- (-0.25,2);
\draw (-1.25,2) arc (180:360:0.5 and -0.2);
\draw (-1.25,0.77) arc (180:360:0.5 and 0.2);
\draw [dashed] (-1.25,0.77) arc (180:360:0.48 and -0.12);
\draw[rotate around={-45:(-0.75,0.75)},red] (-0.75,0.75) ellipse (7pt and 19pt);
\filldraw[black] (0,2) circle (1pt) node[anchor=west]{$x$};
\draw (0.2,0.8) node[label=1:$\mathcal{W}(x)$] (1){};
\draw (-1.03,0.74) node[label=1:$\mathcal{M}(\mathbf{x})$] (2){};
\draw (-2,1) node[label=1:\textcolor{red}{$\mathcal{N}(x)$}] (3){};
\node (4) at (-1,0.4) {};
\draw [->,red] (-1.8,0.9) to [out=-90,in=180] (4.north);
\draw (-0.9,2) node[label=1:$\mathcal{D}$] (4){};
\end{scope}
}
\end{tikzpicture}
    \caption{The past lightcone $\mathcal{C}(x)$ of the field point $x$, where $\mathcal{D}$ is the world tube traced by a codimension-1 sphere of radius $\mathcal{R}$. $\mathcal{C}(x)$ is split into the near zone $\mathcal{N}(x)$ (which lies on the surface of the lightcone and is contained within $\mathcal{D}$) and the wave zone $\mathcal{W}(x)$. The constant-time surface $\mathcal{M}(x)$ is the relevant integration region in the near zone. Figure inspired by illustrations in Chapter 6 of \cite{poisson_will_2014}.}
    \label{fig:lightcone}
    \end{figure}
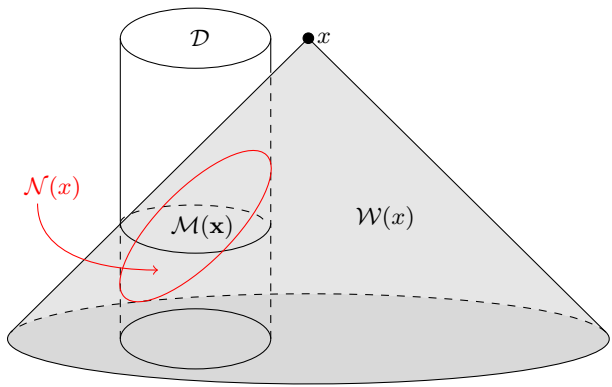
\end{center}

\subsection{Symmetry-breaking solution}
Once the GR solution is safely in hand, we can turn our attention to the first order, i.e. the symmetry-breaking solution; here, derivatives of the full GR solution make up the source term, essentially replacing $\tau^{\mu\nu}$, up to $1$PN order, since the formalism we employ does not allow us to go higher.

\begin{figure}[h]
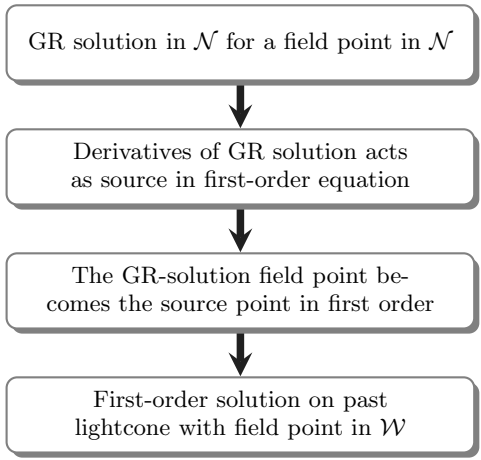

    \begin{center}
       \smartdiagramset{back arrow disabled=true, text width=6cm, uniform color list=white for 4 items,uniform arrow color=true, arrow color=gray!25!black} 
       \smartdiagram[flow diagram:vertical]{{GR solution in $\mathcal{N}$ for a field point in $\mathcal{N}$}, {Derivatives of GR solution acts as source in first-order equation}, The GR-solution field point becomes the source point in first order, First-order solution on past lightcone with field point in $\mathcal{W}$}
    \end{center}
    \caption{The solution-generating algorithm used when evaluating Eq.~(\ref{eq:h1}). Similar logic applies to the wave-zone solutions, but there we will have an extra contribution from the near zone \`{a} la Eq.~(\ref{eq:firstorder}).}
    \label{fig:flow}
\end{figure}

The formal solution for $\bar{h}^{(1)\mu\nu}$ reads
\begin{equation}\label{eq:h1}
    \begin{aligned}
        \bar{h}^{(1)\mu\nu}(x) =& -\frac{1}{2\pi}\int_{\mathcal{N}_\mathcal{W}(x)} d^3x^\prime \frac{\bar{M}^{\mu\nu\rho\sigma}\bar{h}^{(0)}_{\rho\sigma}(\tau,\mathbf{x^\prime})}{|\mathbf{x}-\mathbf{x}^\prime|} \\&- \frac{1}{2\pi}\int_{\mathcal{W}_\mathcal{W}(x)} d^3x^\prime \frac{\bar{M}^{\mu\nu\rho\sigma}\bar{h}^{(0)}_{\rho\sigma}(\tau,\mathbf{x^\prime})}{|\mathbf{x}-\mathbf{x}^\prime|},
    \end{aligned}
\end{equation}
where the numerical prefactors are contained inside $\bar{h}^{(0)}_{\mu\nu}$.
The second integral term is evaluated in the wave zone (with the field point also in the wave zone), where we need to be careful with the field point, since this integral will receive contributions from the GR solution in the near zone as well as the wave zone. We can write the second term in Eq.~(\ref{eq:h1}) as (suppressing some notation)
\begin{equation}\label{eq:wwint}
    \int_{\mathcal{W}_\mathcal{W}(x)} d^3x^\prime \frac{\bar{M}^{\mu\nu\rho\sigma}\left[(\bar{h}^{(0)}_{\rho\sigma}(x^\prime))_{\mathcal{N}_\mathcal{W}}+(\bar{h}^{(0)}_{\rho\sigma}(x^\prime))_{\mathcal{W}_\mathcal{W}}\right]}{|\mathbf{x}-\mathbf{x}^\prime|},
\end{equation}
where we now have two terms with wave-zone field points; in the first-order solution, these will be the source points (see also Figure~\ref{fig:flow}). The second term $(\bar{h}^{(0)}_{\rho\sigma}(x^\prime))_{\mathcal{W}_\mathcal{W}}$, the GR solution in $\mathcal{W}$ with a $\mathcal{W}$ field point, is a $1.5$PN term (a tail effect). This can be seen through explicit evaluation of the GR potential in $\mathcal{W}$, the result of which is proportional to $c^{-3}$ \cite{poisson_will_2014}.
We can therefore write the integral (\ref{eq:h1}) as
\begin{equation}\label{eq:firstorder}
    \begin{aligned}
        \bar{h}^{(1)\mu\nu}(x) =& -\frac{1}{2\pi}\int_{\mathcal{N}_\mathcal{W}(x)}d^3x^\prime \frac{\bar{M}^{\mu\nu\rho\sigma}(\bar{h}^{(0)}_{\rho\sigma}(\tau,\mathbf{x^\prime}))_{\mathcal{N}_\mathcal{N}}}{|\mathbf{x}-\mathbf{x}^\prime|} \\ -& \frac{1}{2\pi}\int_{\mathcal{W}_\mathcal{W}(x)}d^3x^\prime \frac{\bar{M}^{\mu\nu\rho\sigma}(\bar{h}^{(0)}_{\rho\sigma}(\tau,\mathbf{x^\prime}))_{\mathcal{N}_\mathcal{W}}}{|\mathbf{x}-\mathbf{x}^\prime|},
    \end{aligned}
\end{equation}
where the numerator of the integrands can be expanded as
\begin{equation}\label{eq:Mhexp}
    \begin{aligned}
        &\bar{M}^{\mu\nu\rho\sigma}(\bar{h}^{(0)}_{\rho\sigma}(\tau,\mathbf{x^\prime}))_{\mathcal{N}_\mathcal{N}} = \bar{M}^{\mu\nu00}(h^{(0)}_{00}(\tau,\mathbf{x^\prime}))_{\mathcal{N}_\mathcal{N}}\\&+2\bar{M}^{\mu\nu0j}(\bar{h}^{(0)}_{0j}(\tau,\mathbf{x^\prime}))_{\mathcal{N}_\mathcal{N}}+\bar{M}^{\mu\nu jk}(\bar{h}^{(0)}_{jk}(\tau,\mathbf{x^\prime}))_{\mathcal{N}_\mathcal{N}},
    \end{aligned}
\end{equation}
and analogously for $\mathcal{N}_\mathcal{W}$, which we write out to make it clear that we will need {\it all} components of the GR solution\footnote{Note that some components of $\bar{h}^{(1)\mu\nu}$ only show up at mass-dimension $d\geq6$, due to the symmetries of $\bar{M}^{\mu\nu\rho\sigma}.$}; these expressions can be highly non-trivial, depending on the source. This first-order source term appears inside integrals and is expressed using the retarded-time $\tau = t - |\mathbf{x}-\mathbf{x}^\prime|/c$, and since the source term is now of the form $\partial\partial\hdots \bar{h}^{(0)}(x)$, the transformation to retarded time has to take place after the explicit evaluation of the derivatives.

The components of Eq.~(\ref{eq:firstorder}) could use some clarification: bearing in mind that we are interested in the first corrections to GR (at the quadrupole level, since we are working at linear order in the metric potentials), and we have already discarded the GR wavezone integral, since the $\mathcal{W}_\mathcal{W}(x)$ piece lies beyond $1$PN order\footnote{So we also avoid tail effects.}. The remaining piece is the integral over the near zone with a wave-zone field point $x$, which we call $\mathcal{N}_\mathcal{W}(x)$. The source of $\bar{h}^{(1)\mu\nu}$ is now comprised of derivatives of the GR solution $\bar{h}^{(0)\mu\nu}$; here, we keep the near-zone solution, but the {\it field point is now in the near zone as well, since the field point in Eq.~(\ref{eq:GRh}) becomes the source point in the integral (\ref{eq:firstorder})}. In general, the source is defined over all space and does not necessarily have compact support. The chart in Figure~\ref{fig:flow} shows the algorithm we use to find solutions.

\section{Post-Newtonian expansion in the near zone}\label{sec:PN}
Using standard tools, we write the GR solution in the near zone in terms of the Post-Newtonian potentials; the solution (to 1PN order accuracy) reads~\cite{poisson_will_2014}
\begin{equation}\label{eq:h0PNsol}
    \begin{aligned}
        \bar{h}^{(0)00} =& \frac{4}{c^2}U +\frac{1}{c^4}\left(7U^2+4\psi-4V+2\frac{\partial^2 X}{\partial t^2}\right) +\mathcal{O}(c^{-5}) \\
        \bar{h}^{(0)0j} =& \frac{4}{c^3}U^j +\mathcal{O}(c^{-5}) \\ 
        \bar{h}^{(0)ij} =& \frac{1}{c^4}\left(4W^{jk}+U^2\delta^{jk}+4\chi^{jk}\right) + \mathcal{O}(c^{-5}),
    \end{aligned}
\end{equation}
where the potentials satisfy the following Poisson-like equations
\begin{equation}\label{eq:poisson}
    \begin{aligned}
        \nabla^2U =& -4\pi G\rho^*, \\
        \nabla^2U^j =& -4\pi G \rho^* v^j, \\
        \nabla^2X =& 2U \\
        \nabla^2\psi =& -4\pi G\rho^*\left(\tfrac{3}{2}v^2-U+\Pi+3p/\rho^*\right) \\
        \nabla^2V =& -4\pi G\rho^*\left(v^2-\tfrac{1}{2}U+3p/\rho^*\right) \\
        \nabla^2W^{jk} =& -4\pi G\left(\rho^* v^iv^j-\tfrac{1}{2}\rho^*Up\delta^{jk}\right) \\
        \nabla^2\chi^{jk} =& -\partial^jU\partial^kU, 
    \end{aligned}
\end{equation}
the solutions for which are displayed in Appendix~\ref{app:PNpots}. It should be noted that this GR solution goes beyond linear order in the metric potentials, which can be seen by the appearance of terms such as $U^2$ and $\chi^{jk}$; we write down these terms for completeness.
In Eq.~(\ref{eq:h0PNsol}), $\rho^*$ represents the conserved energy density, which to $\mathcal{O}(c^{-4})$ reads $\rho^*=[1-(v^2/2+3U)/c^2]\rho$. The quantities $X$, $V$, $W^{jk}$, and $\chi^{jk}$ are known as superpotentials, since they are defined using the potential $U$. Therefore, we can view the quantity $\bar{M}^{\mu\nu\rho\sigma} \bar{h}^{(0)}_{\rho\sigma}$ as a combination of a superduper potential and a superlative potential\footnote{In the language adopted in the book \cite{poisson_will_2014}.}. 

Now that we have assembled the GR solution in the near zone to the correct PN order, we turn to the source term for first order, which we can write in a schematic way as
\begin{equation}\label{eq:Mhschem}
    \bar{M}^{\mu\nu\rho\sigma}\bar{h}^{(0)}_{\rho\sigma} = \partial \, \partial \, \bar{h}^{(0)} + \partial \, \partial \, \partial \, \bar{h}^{(0)} + \hdots,
\end{equation}
where the solution space already satisfies the generalised first-order gauge condition (\ref{eq:gauge}).
Now, we note that in the near zone, spatial derivatives $\partial_i$ leave the PN order unaffected, whereas temporal derivatives {\it do not}; indeed, since we can write them as

\begin{equation}
    \partial_0 \equiv \frac{\partial}{\partial ct} = \frac{1}{c}\partial_t,
\end{equation}
we see that each consecutive time derivative in (\ref{eq:Mhschem}) adds $\tfrac{1}{2}$ Post-Newtonian order. 
Looking again at the GR solution as
\begin{equation}
    \bar{h}^{(0)00} \sim \frac{4}{c^2}U +\frac{1}{c^4}(\hdots), 
\end{equation}
we see that the first term can be acted on by up to two time derivatives, whilst the second term can only be acted on with spatial derivatives, if we want to keep terms up to $1$PN\footnote{This will be further complicated by the $\mathcal{N}_\mathcal{W}$ integration techniques presented in Section~\ref{sec:sol}, which contain extra factors of $c^{-1}$.}. We also have
\begin{equation}
    \bar{h}^{(0)0j} \sim \frac{4}{c^3}U^j, \quad \bar{h}^{(0)jk} = \frac{1}{c^4}(\hdots),
\end{equation}
so $\bar{h}^{(0)0j}$ can take one time derivative, and $\bar{h}^{(0)jk}$ can only take spatial derivatives. This will somewhat simplify the following calculations in the near zone. Due to the use of the retarded time in the wave zone, we will not have the freedom in spatial derivatives as in the near zone, which is shown in Eq.~(\ref{eq:derivatives}).

\section{Wavezone-wavezone contribution}\label{sec:grsolint}
At the GR level, the $\mathcal{W}_{\mathcal{W}}$ contribution appears at $1.5$PN \cite{poisson_will_2014}, and we will therefore not consider it; however, the $\mathcal{N}_{\mathcal{W}}$ term integrated over $\mathcal{W}_{\mathcal{W}}$ survives. The $\mathcal{N}_{\mathcal{N}}$ contribution integrated over $\mathcal{N}_{\mathcal{W}}$ will be our main focus for the purposes of the toy solution in Section~\ref{sec:sol}, but for certain components of the SME coefficients, the $\mathcal{W}_\mathcal{W}$ contribution will introduce terms at the same Post-Newtonian order. From Eq.~(\ref{eq:firstorder}) we have that
\begin{equation}\label{eq:firstorderWW}
        \bar{h}^{(1)\mu\nu}(x) \supseteq -\frac{1}{2\pi}\int_{\mathcal{W}_\mathcal{W}(x)}d^3x^\prime \frac{\bar{M}^{\mu\nu\rho\sigma}(\bar{h}^{(0)}_{\rho\sigma}(\tau,\mathbf{x^\prime}))_{\mathcal{N}_\mathcal{W}}}{|\mathbf{x}-\mathbf{x}^\prime|}.
\end{equation}
The GR solution in $\mathcal{N}_{\mathcal{W}}$ reads (up to $1$PN accuracy)
\begin{equation}\label{eq:GRNW}
    \begin{aligned}
        \bar{h}^{(0)00} =& \frac{4G}{c^2}\left[\frac{M}{r}+\frac{1}{2}\partial_j\partial_k\frac{\mathcal{I}^{jk}}{r}\right], \\ \bar{h}^{(0)0j} =& \frac{4G}{c^3}\left[-\frac{1}{2}\frac{(\mathbf{n}\times\mathbf{J})^j}{r^2}-\frac{1}{2}\partial_k\frac{\dot{\mathcal{I}}^{jk}}{r}\right] \\
        \bar{h}^{(0)jk} =& \frac{4G}{c^4}\left[\frac{1}{2}\frac{\ddot{\mathcal{I}}^{jk}}{r}\right],
    \end{aligned}
\end{equation}
where $M$ is the total ADM (Arnowitt-Deser-Misner) mass of the system (which agrees with the difference between the monopole moment and the near-zone mass to order $c^{-4}$), $J$ is the total angular momentum (with agrees with the near-zone angular momentum to $c^{-2}$), $\mathcal{I}^{jk}$ is the mass quadrupole moment, $\mathbf{n}$ is a normal vector pointing from the source point to the field point (from source to detector), and $r=|\mathbf{x}-\mathbf{x^\prime}|$ is the distance between the same. Since $x$ is in $\mathcal{W}$, $r$ can not be considered a small quantity. Overdots indicate derivatives w.r.t. retarded time $\tau$. This expression can be simplified in the far-away wavezone (when $r$ is much larger than the characteristic wavelength of the source), and this is what is generally done for GR solutions; however, since these potentials make up our source for the first-order equations and will be integrated over all of $\mathcal{W}$, we must keep their full form. 

The solution (\ref{eq:GRNW}) will be differentiated partially a number of times in the same manner as in the near zone, where spatial derivatives generate more normal vectors when acting on $r^{-1}$. When counting Post-Newtonian orders in this region, we note that the total mass $M$ contains a $0$PN component, since it is defined as
\begin{equation}
    M = \int d^3x \rho^*\left[1-\frac{1}{c^2}\left(\frac{1}{2}v^2-\frac{1}{2}U+\Pi\right)\right] + \mathcal{O}(c^{-4}),
\end{equation}
where $\Pi$ is the specific internal energy; the PN contribution here is the velocity-order expansion in $v/c$. The retarded-time derivatives of the mass quadrupole moment can be evaluated using standard formulae\footnote{See Box 7.7 in \cite{poisson_will_2014}.}. When evaluating the partial derivatives present in $\bar{M}^{\mu\nu\rho\sigma}$ we note that in contrast to the $\mathcal{N}_\mathcal{N}$ derivatives, {\it both} temporal and spatial derivatives increase the Post-Newtonian order here. Since the source is now retarded, acting with a non-retarded derivatives (on a generic retarded function $f(\tau)$) results in
\begin{equation}\label{eq:derivatives}
        \frac{1}{c}\partial_t f(\tau) = \frac{1}{c}\partial_\tau f(\tau), \quad \partial_i f(\tau) = -\frac{1}{c}\frac{n_i}{r} \partial_\tau f(\tau),
\end{equation}
and we conclude that we no longer have the same freedom to add spatial derivatives; any derivatives (both temporal and spatial derivatives) will add a corresponding factor of $c^{-1}$. There is, however, freedom remaining in the factors of $r^{-1}$ which appear in Eq.~(\ref{eq:GRNW}), which when acted on with spatial derivatives will generate normal vectors $n_i$ without increasing the Post-Newtonian order.

Now that we have obtained the derivatives of the GR source and applied the corresponding SME coefficients, we note that when integrating over $\mathcal{W}$ with the field point in $\mathcal{W}$, the general solution to Eq~(\ref{eq:firstorderWW}) reads \cite{poisson_will_2014}
\begin{equation}\label{eq:firstorderWWsol}
    \begin{aligned}
        \bar{h}^{(1) \mu\nu}(x) \supseteq -\frac{1}{2\pi}\frac{n^{\langle L\rangle}}{r}&\Bigg[\int_0^\mathcal{R}ds f(\tau-2s/c)A(s,r) \\ &+\int_\mathcal{R}^\infty ds f(\tau-2s/c)B(s,r)\Bigg],
    \end{aligned}
\end{equation}
where $\mathcal{R}$ denotes the boundary between $\mathcal{N}$ and $\mathcal{W}$, and $A,B$ are integrals of Legendre polynomials. The symbol $n^{\langle L \rangle}$ is a symmetric and trace-free (STF) product of radial vectors; for example, $n^{\langle j k \rangle}=n^jn^k-\tfrac{1}{3}\delta^{jk}$. The expressions depend on the specific form of the source function, in our case $\bar{M}\bar{h}^{(0)}$, where we are forced to restrict our attention to sources of the form (dropping some notation)
\begin{equation}\label{eq:f}
    f(\tau) = 4\pi \left(\frac{n^{\langle L \rangle}}{r^n}\right)^{-1} \bar{M}^{\mu\nu\rho\sigma}\bar{h}^{(0)}_{\rho\sigma},
\end{equation}
where $n^{\langle L \rangle}$ can be written as a combination of spherical harmonics and integrated trivially. Therefore, when constructing $f(\tau)$, it will be necessary to write it as a sum of subsources, each satisfying the prescription in Eq.~(\ref{eq:f}) with the correct identification of $L$ and $n$. We will not examine these contributions further, and instead leave it for future work, which includes an exhaustive set of solutions \cite{solpaper}.

\section{The quadrupole formula}\label{sec:quadrupole}
In this paper, we are interested in the lowest-order corrections to GR, which will appear as corrections to the quadrupole formula (or at least at the quadupole order), a contribution at 1PN order. In GR, the quadrupole tensor appears in the metric potentials as
\begin{equation}
    \begin{aligned}
        \bar{h}^{00}\supseteq \partial_j\partial_k\left(\frac{\mathcal{I}^{jk}}{r}\right),  \bar{h}^{0j}\supseteq \partial_k\left(\frac{\dot{\mathcal{I}}^{jk}}{r}\right), \bar{h}^{jk}\supseteq \frac{\ddot{\mathcal{I}}^{jk}}{r}
    \end{aligned}
\end{equation}
up to constant factors and with some corrections at higher order in derivatives, where $\mathcal{I}^{jk}$ is the mass quadrupole moment. This can also be seen from the harmonic gauge condition at the GR level\footnote{The authors of \cite{Bailey:2006fd} find a gauge equivalent to the harmonic gauge up to $1.5$PN order for the PN metric $g_{\mu\nu}$ using $g_{\mu\nu}=\eta_{\mu\nu}+h_{\mu\nu}$ and a different representation of the SME coefficients.}, $\partial_\mu \bar{h}^{(0)\mu\nu}=0$. In our case,
we will compute it from the spatial components $\bar{h}^{(1)jk}$. Looking at the first-order source term needed, it is the $\mu=j, \nu=k$ component of Eq.~(\ref{eq:Mhexp}) which reads (suppressing some notation)
\begin{equation}\label{Mhjk}
    \bar{M}^{jk00}\bar{h}^{(0)}_{00}+2\bar{M}^{jkm0}\bar{h}^{(0)}_{m0}+\bar{M}^{jkmn}\bar{h}^{(0)}_{mn},
\end{equation}
which can be partially simplified using the index symmetries of $\hat{s}^{\mu\rho\nu\sigma}$, $\hat{q}^{\mu\rho\nu\sigma}$, and $\hat{k}^{\mu\rho\nu\sigma}$\footnote{Note also that $\bar{M}^{\mu\nu\rho\sigma}\bar{h}^{(0)}_{\rho\sigma} = M^{\mu\nu\rho\sigma}h^{(0)}_{\rho\sigma}$}.
Armed with this knowledge, we can now form the source for the modifications to the quadrupole formula, using Eq.~(\ref{Mhjk}), which contains $\geq2$ derivatives of the GR solution. Note that we still have only made some very general assumptions about the properties of the source; indeed, beyond the existence of a near zone, our results are valid for any source up to this point (but we will have to limit the set of sources to those which can be described by the linear potentials). 

For a generic source with a wave-zone field point, we can write the linear solution to the spatial metric potential as
\begin{equation}
    \bar{h}^{jk} = \frac{G}{c^4R}A^{jk},
\end{equation}
and in GR, we have that $A^{jk}=2\ddot{\mathcal{I}}^{jk}$, where an overdot denotes a derivative w.r.t retarded time $\tau$, and the physical piece of $\bar{h}^{jk}$ is the transverse-traceless part $\bar{h}^{jk}_{\rm TT}$. When spacetime symmetries are broken, we generally expect additional polarisation modes to appear, up to a total of six, i.e. four extra modes may appear in the spectrum of gravitational waves\footnote{In \cite{ONeal-Ault:2020ebv} one of us found one extra dynamical degree of freedom using the Hamiltonian formulation in a simple explicit-breaking scenario, which indicates the presence of extra modes.}; however, these extra modes will be suppressed proportional to the coefficients for spacetime-symmetry breaking, which are very small\footnote{We also expect the transverse-traceless part to receive modifications.}. This can be seen more easily in momentum space, where the dispersion relation for the helicity components $\bar{h}_{\pm 2}$ can be solved for in an order-by-order fashion. This yields a result which is neither transverse nor traceless, but the dominant effects can be captured by the TT piece of the dispersion relations \cite{Mewes:2019dhj}. We note, however, that using the gauge invariant hatted operators $\hat{s}^{\mu\nu\rho\sigma}$, $\hat{q}^{\mu\nu\rho\sigma}$, and $\hat{k}^{\mu\nu\rho\sigma}$ as we do in this paper, only two polarisation modes appear. Other choices of coefficients, for example the ``barred'' coefficients\footnote{Which are related to the hatted operators through Eq.~(\ref{eq:hatbar}).} used in \cite{Bailey2023} allows for five out of six polarisation modes.

In order to obtain the corrections to the quadrupole formula, we will use the spatial part of the solution $\bar{h}^{(1)jk}$, and we will first need to evaluate some rather cumbersome derivatives of the metric potentials (\ref{eq:h0PNsol}). These can be made more tractable by introducing some simplifications (for example point-particle approximations), but it should be understood that these derivatives must be taken {\it before} the GR solution $\bar{h}^{{(0)}jk}$ is plugged in as a source on the right-hand side of Eq.~(\ref{eq:firstorder}) (where it is evaluated at the retarded time $\tau$ rather than $t$). We will see an exception to this rule when introducing point particles in the next section.

We now turn back to Eq.~(\ref{eq:firstorder}) and spend some time preparing to solve it in $\mathcal{N}_\mathcal{W}$, i.e. in the near zone with a wave-zone field point. For a generic wave equation $\Box \psi^{\mu\nu}=-4\pi \xi^{\mu\nu}$, the solution takes the form of a multipole expansion and reads (See Chapter 6 of \cite{poisson_will_2014}).
\begin{equation}\label{eq:NWsol}
    \psi^{\mu\nu}(x) = \sum_{\ell=0}^\infty\frac{(-1)^\ell}{\ell!}\partial_L\left[\frac{1}{r}\int_\mathcal{M}d^3x^\prime\xi^{\mu\nu}(\tau,\mathbf{x^\prime})x^{\prime L}\right],
\end{equation}
where $L$ is a multi index defined such that $x^L = x^{j_1}x^{j_2}\hdots x^{j_\ell}$ ($j_1, j_2$ etc are spatial indices), with contractions of $L$ following standard Einstein summation. In Eq.~(\ref{eq:NWsol}), the integration region $\mathcal{M}$ is a surface of constant time bounded externally by a sphere $r'=\mathcal{R}$ which is defined as the edge of the near zone\footnote{The expression (\ref{eq:NWsol}) is valid for any field point $x$ in the wave zone, but can be further simplified for $r\to \infty$, a simplification which we adopt in Section \ref{sec:sol}.}. A depiction of the relevant integration regions is shown in Figure~\ref{fig:lightcone}.

In order to obtain the quadrupole-order expression, we truncate the series in Eq.~(\ref{eq:NWsol}) at $\ell=2$, after which the $\ell=2$ GR solution of the spatial components can be written as
\begin{equation}
    \begin{aligned}
        \psi^{jk} =& \frac{2G}{c^4}\frac{\ddot{\mathcal{I}}^{jk}}{r},\quad
        \mathcal{I}^{jk}(\tau) = \int d^3x \rho^* x^jx^k
    \end{aligned}
\end{equation}
to $1$PN order. As a consequence of this, our final solution will naturally contain derivatives of the GR quadrupole expression, but since we now have a non-standard source, the monopole ($\ell=0$) and dipole ($\ell=1$) may no longer vanish as they do in GR.

\section{Toy solution}\label{sec:sol}
Here, we show some typical computations involved in finding the explicit solutions. To make this example more tractable, we introduce the following simplifications
\begin{enumerate}
    \item Consider only mass-dimension $d=4$ SME coefficients;
    \item Keep only certain components of the SME coefficients in order to keep only terms linear in the potentials;
    \item Discard contributions from the $\mathcal{N}_{\mathcal{W}}$ integration of the GR solution (\ref{eq:wwint}).
\end{enumerate}
We note here that the $d=4$ truncation in the first simplification is known as the minimal SME and is a common choice in the literature, whereas simplification $2$ and $3$ are introduced purely to produce tractable expressions in the toy solution; these may not be admissible for real phenomenological studies.

We can now carry out the below computations by hand and write down a small number of terms as the final solution; the complete solution will be significantly more complicated, but will be constructed of the same types of terms.

We consider the simple case of a system of point particles and mass-dimension $d=4$ in SME coefficients, so that only $\hat{s}^{\mu\rho\sigma\nu}$ contributes in the first-order solution (\ref{eq:firstorder}). In this case we will have
\begin{equation}
    M^{\mu\nu\rho\sigma} = -\tfrac{1}{4}\left[\hat{s}^{\mu\rho\nu\sigma}+\hat{s}^{\mu\sigma\nu\rho}\right],
\end{equation}
and the spatial part of the source term reads
\begin{equation}\label{eq:exampleM}
    \bar{M}^{jk00}\bar{h}^{(0)}_{00}+2\bar{M}^{jkm0}\bar{h}^{(0)}_{m0}+\bar{M}^{jkmn}\bar{h}^{(0)}_{mn}.
\end{equation}
Looking at $\hat{s}^{\mu\rho\nu\sigma}=s^{(4)\mu\rho\alpha\nu\sigma\beta}\partial_\alpha\partial_\beta$, we simplify further by demanding that only the middle term of the source in Eq.~(\ref{eq:exampleM}) contributes, which can be achieved by only letting a certain subset of coefficient component be non-zero. We call that subset of terms $\hat{\tilde{\bar{s}}}^{jk\alpha\beta}$, after we can write the middle term as
\begin{equation}
    2\bar{M}^{jkm0}\bar{h}^{(0)}_{m0} = -\tfrac{1}{2}\tilde{\bar{s}}^{jk\alpha\beta}\partial_\alpha\partial_\beta \bar{h}^{(0)}_{m0}. 
\end{equation}
Counting now the PN orders of this term, we see from the $0j$ component of Eq.~(\ref{eq:h0PNsol}) that we are allowed exactly one time derivative in order for the resulting expression to be a $1$PN contribution to the solutions, so we set $\alpha=0$, $\beta=i$ and find that the final source term reads
\begin{equation}
    2\bar{M}^{jkm0}\bar{h}^{(0)}_{m0} = -\tfrac{1}{2}\tilde{\bar{s}}^{jkmi}\tfrac{1}{c}\partial_t\partial_i \bar{h}^{(0)}_{m0}.
\end{equation}
We now turn to the GR solution $\bar{h}^{(0)}_{m0}$ in Eq.~(\ref{eq:h0PNsol}), which consists of the potential $U_j(x)$. For a system of isolated bodies, we can write the potential for body $A$ as
\begin{equation}
    U_j(x) = U_j^A(x) + U_j^{\neg A}(x),
\end{equation}
where
\begin{equation}
    U_j^A(x)=\sum_A\frac{Gm_Av^A_j}{|\mathbf{x}-\mathbf{r_A}(t)|}, U_j^{\neg A}(x)=\sum_{B\neq A}\frac{Gm_Bv^B_j}{|\mathbf{x}-\mathbf{r_B}(t)|}
\end{equation}
and because of the wide separation between the bodies\footnote{Which is certainly true for the case of point particles during the inspiral phase.}, we can write the external potential as a Taylor expansion
\begin{equation}
    \begin{aligned}
        U^{\neg A}_j(t,\mathbf{x}) &= U^{\neg A}_j(t,\mathbf{r}_A)\\&+|\mathbf{x}-\mathbf{r}_A|^k\left(\partial_k U_j^{\neg A}(t,\mathbf{x})\right)\large|_{\mathbf{x}\to\mathbf{r}_A}+\hdots,
    \end{aligned}
\end{equation}
where we neglect terms beyond linear order and take the derivative before setting $\mathbf{x}\to\mathbf{r}_A$. After taking the derivative and simplifying terms, the external potential reads
\begin{equation}\label{eq:unota}
    U^{\neg A}_j(t,\mathbf{x})=G\sum_{B\neq A}\frac{m_Bv^B_j}{r_{AB}}\left[1-(\mathbf{x}-\mathbf{r}_A)^k\frac{n^{AB}_k}{r_{AB}}+\hdots\right],
\end{equation}
where we have introduced the notation $\mathbf{n}^{AB}=\mathbf{r^{AB}}/r^{AB}$ and $\mathbf{r}^{AB}=\mathbf{r}^A-\mathbf{r}^B$. 

Looking now at the first-order source term, which appears in our correction to the first-order potentials as $\int_M d^3x^\prime \bar{M}^{\mu\nu\rho\sigma}\bar{h}^{(0)}_{\rho\sigma}/|(\mathbf{x'}-\mathbf{r}(t))|$, we note that since the derivatives in $\bar{M}^{\mu\nu\rho\sigma}$ appear under an integral sign, we must treat them as {\it distributional derivatives} rather than simple partials, precisely because there are singularities present. The need to treat derivatives in this way normally appears at the 2PN order, but since our case involves a type of bootstrapping of the GR solution, we can expect that such complications will appear at lower order. When taking derivatives of non-compact potentials in $D$ dimensions, non-compact products of Dirac distributions appear, which can be counteracted by considering the Schwarz distributional derivative \cite{schwartz,Blanchet:2000nu,Marchand:2020fpt} which we denote $D_i[\cdot]$ and can be obtained from the generalised Gel’fand-Shilov formula~\cite{gelshi}
\begin{equation}\label{eq:distrder}
\begin{aligned}
    D_i[F] = \sum_{\ell\geq0}&\frac{(-1)^\ell}{\ell!}\partial_L\delta^{(3)}(\mathbf{x}-\mathbf{r_1}(t))\int d\Omega_1 (n_i)_1 \, n_1^L \, (f_{-2-\ell})_1 \\ +& 1 \leftrightarrow 2 = -\underset{1}{D_i}[F]-\underset{2}{D_i}[F],
    \end{aligned}
\end{equation}
where the minus signs appear due to the potential $F$ only depending on the velocities through the distance to the field point\footnote{see Theorem 4 and Section 9 in \cite{Blanchet:2000nu}. The numbers under the terms refers to the different particles in the two-body problem.}, $D_i[F]+\underset{1}{D_i}[F]+\underset{2}{D_i}[F]=0$.
A spatial partial derivative of a singular potential $F$ under an integral is therefore generalised to read $\partial_iF \mapsto (\partial_iF)_{\rm can}+D_i[F]$, where $(\partial_i)_{\rm can}$ is the canonical basis of the tangent space and $D_i[F]$ is the distributional correction. This definition can be generalised to arbitrary derivative order, see for example \cite{gelshi}. Time derivatives are similarly defined as $\partial_t F\mapsto(\partial_t F)_{\rm can}+D_t[F]$, where the distributional correction is proportional to the velocity of the source contracted with $D_i[F]$. We will also need to generalise the case where spatial and temporal partial derivatives are mixed, in which case we have $\partial_t\partial_{i}F=(\partial_t\partial_i)_{\rm can}+D_t[\partial_iF]+\partial_tD_i[F]$, which can also be generalised as necessary. We note here that distributional derivatives do not a priori commute, but it can be shown that this ambiguity only becomes important at $4PN$ order \cite{Marchand:2020fpt}.

We note here that the derivative in the Taylor expansion does not need to be considered as distributional, since it is an internal pure derivative and not part of the physical solution. When we compute the two partial derivatives contained in the {\it source} $M h$, the distributional parts will be important. Nevertheless, we will also need the canonical partial derivatives of the above potentials, but we note that $U^{\neg A}_j(t,\mathbf{x})$ lacks singularities and we therefore do not need to introduce the distributional derivative correction or the Hadamard regularisation for this term. The two (canonical) partial derivatives of $U^{\neg A}_j(t,\mathbf{x})$ can be trivially computed. The situation is more involved for $U^A_j(t,\mathbf{x})$ due to the singularities present when $\mathbf{x}\to\mathbf{r}_A$; here, we compute the mixed distributional derivative and find
\begin{equation}\label{eq:sourceexpdistrder}
    \begin{aligned}
        \partial_t\partial_i U^A_j(t,\mathbf{x}) = (\partial_t\partial_i U^A_j(t,\mathbf{x}))_{\rm can} &+ D_t[\partial_i U^A_j(t,\mathbf{x})] \\&+\partial_tD_i[U^A_j(t,\mathbf{x})],
    \end{aligned}
\end{equation}
From Eq.~(\ref{eq:distrder}) we see that $D_i[U^A_j(t,\mathbf{x})]$ does not generate a distributional part and thus vanishes, but because of the spatial derivative, $D_t[\partial_i U^A_j(t,\mathbf{x})]$ remains; after some computation, we see that it equates to
\begin{equation}\label{eq:distrdert}
    D_t[\partial_i U^A_j(t,\mathbf{x})] = -\frac{2G}{3}\sum_A m_Av_i^A(t)v_j^A(t)\delta^{(3)}(\mathbf{x}-\mathbf{r_A}(t)).
\end{equation}
Now that we have gathered all the results we need for the source term $\partial_t\partial_i U_j(x)$, we are left with the task of carrying out the integration in Eq.~(\ref{eq:firstorder}). As we are integrating in the near zone with a wave-zone field point ($\mathcal{N}_\mathcal{W}(x)$), the problem reduces to evaluating Eq.~(\ref{eq:NWsol}) over the constant-time hypersurface $\mathcal{M}(\mathbf{x})$ as depicted in Figure~\ref{fig:lightcone}, when the source term is evaluated at the retarded time, $(\partial_t\partial_i \bar{h}^{(0)}_{m0})|_{t\to\tau}$. First, we evaluate Eq.~(\ref{eq:NWsol}) in the {\it far-away wavezone}, i.e. for the limit $r\to\infty$, and we retain only the term linear in $1/r$. We also note that since the only dependence on the unprimed coordinates $\mathbf{x}$ in $\tau^{\mu\nu}$ is through the retarded time, and we can write $\partial_L\tau^{\mu\nu} = (-1)^\ell c^{-\ell} \partial_\tau^\ell \tau^{\mu\nu} n_L +\hdots$ \cite{poisson_will_2014}, after which Eq.~(\ref{eq:NWsol}) becomes
\begin{equation}\label{eq:NW}
    \bar{h}^{(1)\mu\nu}(x) = -\frac{1}{2\pi r}\sum_{\ell=0}^\infty \frac{n_L}{\ell!c^\ell}\left(\frac{d}{d\tau}\right)^\ell\int_{\mathcal{M}} d^3x^\prime\tau^{(0) \mu\nu}(\tau,\mathbf{x^\prime}) x^{\prime L}
\end{equation}
to first order in $1/r$, where we have added the superscript $(0)$ on the source function $\tau^{(0)\mu\nu}$ to make it clear that it consists of the GR-level solution $\bar{h}^{(0)\mu\nu}$ and derivatives of the same. From this formula, we can read off that the expected quadrupolar expression corresponds to $\ell=2$, and for a simple mass distribution in GR, both the monopolar and dipolar contributions vanish, which may no longer be true in our case. We also notice the following: the quadrupolar term $\ell=2$ appears with an accompanying factor $c^{-2}$, which together with the $c^{-4}$ already present within our source, we have in essence a $2$PN contribution from the ``quadrupole'' term. If we had chosen two spatial derivatives in this example, the same term would contribute at $1.5$PN. We now look at the monopole ($\ell=0$) and dipole ($\ell=1$), which should appear as $1$ and $1.5$PN contributions, respectively. 
The spatial part of the source term in (\ref{eq:NW}) reads explicitly
\begin{equation}
    \tau^{(0) jk}(\tau,\mathbf{x^\prime}) = \frac{2}{c^4}\widetilde{\bar{s}}^{jkmi}\left(\partial_t\partial_i U_m(t,\mathbf{x})\right){\Large|}_{\substack{t\to\tau\\\mathbf{x}\to\mathbf{x^\prime}}},
\end{equation}
and where the derivatives include distributional pieces ($\partial f \equiv (\partial f)_{\rm can}+D[f]$), since $\tau^{(0) jk}$ appears under an integral. In principle, more terms contribute at the same PN order in the above equation, but due to our very specific choice of the non-zero SME coefficients, only one term survives.

\subsection{Monopole term}
Looking again at the spatial components, the $\ell=0$ term reads
\begin{equation}
    \bar{h}^{(1)jk}(x) \supseteq \bar{h}^{(1)jk}_{\ell =0}(x) = -\frac{1}{2\pi r}\int_\mathcal{M}d^3x^\prime \tau^{(0) jk}(\tau,\mathbf{x^\prime}),
\end{equation}
which does not come with extra factors of $c^{-1}$ except for those which may or may not be present in $\tau^{(0)jk}$. It is now important to count the Post-Newtonian orders: for the purposes of this example, we chose a source containing $\partial_0\partial_i$, which contributes $c^{-1}$ to the $c^{-3}$ already present in the ($0j$) component of the GR solution. Therefore, since we wish to stay at $1$PN order, we must truncate the sum over $\ell$ in Eq.~(\ref{eq:NW}) to $\ell=0$, i.e. the {\it symmetry-breaking contribution will be a monopole term}\footnote{Although with two free indices since we are focusing on the spatial components of $\bar{h}^{(1)\mu\nu}$, so it is only a monopole term in the sense that $L=0$ in Eq.~(\ref{eq:NW}).}. If we had chosen the source such that only spatial derivatives appeared, we would have also retained the dipole term at $1$PN order.
Now, using the results obtained from Eqs.~(\ref{eq:sourceexpdistrder}) and (\ref{eq:distrdert}) we see that only three of the terms are divergent in the near zone. We have $\partial_t\partial_i (U^A_j(\mathbf{x})+U^{\neg A}_j(\mathbf{x}))$, the pieces of which are all presented in the previous subsection, with the divergent piece being $(\partial_t\partial_iU^A_j(t,\mathbf{x}))_{\rm can}$, which will contribute to the solution as
\begin{equation}
    \begin{aligned}
        (\bar{h}^{(1)jk})_{\rm div}& = -\frac{G}{\pi c^4 r}\widetilde{s}^{jkmi}\sum_A\int_\mathcal{M}d^3x^\prime \frac{m_A a_m^An^\prime_{A,i}}{|\mathbf{x^\prime}-\mathbf{r_A}(\tau)|^2}\\&-\frac{v_i^Av_m^A}{|\mathbf{x^\prime}-\mathbf{r_A}(\tau)|^3}+3\frac{n_{A,i}^\prime v_m^A(\mathbf{n^\prime_A}\cdot\mathbf{v_A})}{|\mathbf{x^\prime}-\mathbf{r_A}(\tau)|^3},
    \end{aligned}
\end{equation}
where $a_m^A$ is the Newtonian acceleration and where the source is now evaluated at the retarded time, so under the integral $\tau^{jk}=\tau^{jk}(\tau,\mathbf{x^\prime})$. First, we notice that the last two terms can be written as a spherical average of an STF tensor (symmetric and trace free \cite{poisson_will_2014}), which integrates to zero under Gauss' theorem; therefore, only the first term contributes.

To compute the integrals in this section we make use of the Hadamard Finite Part (Partie Finie) regularisation procedure, which has been widely used in the literature up to $3$PN; for details, see for example \cite{Blanchet:2000nu, Blanchet:2013haa}. At higher PN orders, true ambiguities arise\footnote{Interestingly, these ambiguities include the possible loss of Lorentz and diffeomorphism invariance \cite{Blanchet:2005tk}.}, and one is forced to resort to dimensional regularisation \cite{Blanchet:2003gy,Blanchet:2005tk}. After some algebra and plugging in the expression for the Newtonian equation of motion, we obtain
\begin{equation}
    (\bar{h}^{(1)jk})_{\rm div}= \frac{4G^2}{3c^4r}\widetilde{\bar{s}}^{jkmi}\frac{m_1m_2n_{i}^{12}n_{m}^{12}}{r_{12}},
\end{equation}
which is a $1$PN term. Here, we note that the Newtonian equation of motion in principle also acquires symmetry-breaking corrections of the form $A_{ij}n^j GM/r^2+\hdots$, where $A_{ij}$ represents the SME coefficients; this will add extra symmetry-breaking terms in the final solution.
The terms left to integrate are now those without singular denominators, i.e. Eq.~(\ref{eq:unota}) and (\ref{eq:distrdert}). First, we note that thanks to the appearance of the delta function from the distributional time derivative, Eq.~(\ref{eq:distrdert}) can be rewritten as
\begin{equation}
    (\bar{h}^{(1)jk})_{\rm distr}=-\frac{4G}{3c^4r}\widetilde{\bar{s}}^{jkmi}\sum_A m_A v^A_iv^A_m.
\end{equation}
We can rewrite the parts of the solution arising from the distributional time derivative and the divergent integral as
\begin{equation}\label{eq:finalsol}
        (\bar{h}^{(1)jk})_{\rm distr}+(\bar{h}^{(1)jk})_{\rm div} = -\frac{4G}{3c^4r}\widetilde{\bar{s}}^{jkmi} \ddot{I}^{\rm GR}_{im}
\end{equation}
where $I^{\rm GR}_{im}$ is the GR quadrupole-moment tensor and dots indicate derivatives w.r.t the retarded time, since this arises from the retarded source under the integral. 

Lastly, we note that the term containing Eq.~(\ref{eq:unota}) will not have distributional contributions, since it is finite everywhere in the near zone. We also note that since it was constructed by means of a Taylor expansion, the expression only holds when $\mathbf{x}-\mathbf{r_A}$ is small, i.e. close to particle $A$, and that it contains a growing piece outside of the near zone, which is not physical; therefore we assume, for the purposes of this toy solution, that Eq.~(\ref{eq:unota}) has compact support inside the near zone. It is also possible to write it as an integral of a gradient, since in the monopole term $U^{\neg A}_m$ is acted on by $\tfrac{1}{c}\widetilde{\bar{s}}^{jkmi}\partial_i\partial_t$; we let the time derivative act and then write the integral as
\begin{equation}
    \int_\mathcal{M} d^3x' \partial_i F^{ijk}(\tau,\mathbf{x^\prime}),
\end{equation}
where $F^{ijk}\equiv \widetilde{\bar{s}}^{jkmi}U^{\neg A}_m$. Now, thanks to the compact-support assumption, this integral vanishes. This leads us to the conclusion that in the point-particle case, the only contribution comes from the singularities themselves\footnote{To some reasonable approximation, a statement which will be refined in \cite{solpaper}.}. The final result at the SME level is therefore the right-hand side of Eq.~(\ref{eq:finalsol}), which is rather elegant from the point of view of observation, since the correction is proportional to the GR quadrupole tensor. To this we should of course add the well-known GR solution for a field point in the wave zone, i.e. {\it not} the expression presented in Eq.~(\ref{eq:h0PNsol}), which was used to construct the first-order source term $\tau^{(0)\mu\nu}$ and is only valid when both the field point and source point are in the near zone. Finally, the solution reads
\begin{equation}\label{eq:samplesolfinal}
    \bar{h}^{(1)jk} \supseteq \bar{h}^{{\rm GR} jk}_{\mathcal{N}_\mathcal{W}}-\frac{4G}{3c^4r}\widetilde{\bar{s}}^{jkmi}\ddot{I}^{\rm GR}_{im}+\mathcal{O}(c^{-5}),
\end{equation}
which is a remarkably simple expression. Here $\bar{h}^{{\rm GR} jk}_{\mathcal{N}_\mathcal{W}}$ is the GR solution in the near zone with a wave-zone field point expressed at 1PN order (which is the quadrupole). 
We note that this is highly simplified due to the simplifications introduced in the beginning of this section; in the complete solution, terms of the form $\bar{M}^{jk00}\bar{h}^{(0)}_{00}+\bar{M}^{jkmn}\bar{h}^{(0)}_{mn}$ will appear, where $\bar{h}^{(0)}_{\mu\nu}$ are those in Eq.~(\ref{eq:h0PNsol}), which will be significantly more complicated given the complete set of terms and the contribution from the $\mathcal{W}_\mathcal{W}$ integration. It should also be pointed out that the GR quadrupole tensor will also contain corrections coming from the SME terms: when deriving the expression (\ref{eq:samplesolfinal}), we substituted in the Newtonian equation of motion and the virial theorem, both of which contain modifications proportional to $\widetilde{\bar{s}}^{jkmi}$ which we do not write out explicitly here. Such corrections show up naturally when spacetime symmetries are broken, and were for example found in the modified precession equations derived in \cite{Bailey:2013oda}.

\section{Discussion and future work}\label{sec:disc}
In this paper we introduced the tools and methodology necessary for studying the effects of spontaneous spacetime-symmetry breaking in the generation stage of gravitational waves. Working with the Landau-Lifshitz formulation of GR (to first order in the metric potential $h$) and the operator formulation of the Standard-Model Extension gravitational sector, we wrote down the modified relaxed Einstein equations to arbitrary order in operator mass dimension $d$, which contains $d-2$ derivatives of the GR potentials $\bar{h}^{(0)\mu\nu}$ (in natural units). By employing an order-by-order solution strategy where the GR solution acts as the source we wrote down the formal solution, which consists of nested Green's functions, since we are solving an inverse d'Alembertian problem. We also solved a simplified toy example: by using the Post-Newtonian expansion in the near-zone of the source for the case of point particles, we were able to regularise the divergent pieces of the integrals using Hadamard regularisation, and in a simple sample solution, we see that the monopolar and dipolar contributions do not in general vanish, which is in contrast to the case of General Relativity. Throughout the paper, we carefully discussed the PN order of the various terms, keeping in mind that we are ultimately interested in the $1$PN corrections to GR. In future work, we will consider higher-order PN corrections, tail effects, as well all subtleties discussed in this paper.

Future space-based gravitational-wave observatories such as LISA can potentially detect signals of the type derived here; indeed, considering the long integration times available for galactic binaries (a minimum of four years, the nominal LISA mission lifetime), the amount of statistics available is going to be considerable. In Eq.~(\ref{eq:samplesolfinal}), the symmetry-breaking contribution is suppressed by $\widetilde{\bar{s}}^{jkmi}$, a linear combination of SME coefficients; these are the coefficients we are interested in constraining, and comparing to other bounds from gravitational waves, we see that SME coefficients are typically constrained at the level of $10^{-14}-10^{-16}$ \cite{Kostelecky:2008ts}. Therefore, it will likely be necessary to combine observations from several galactic binaries and to use long integration times. Also, we expect that a more complete solution will contain more exotic polarisation modes and other standard terms, which can be given tight constraints.

The methods we have outlined in this paper can be applied to any source where the Post-Newtonian expansion is valid, for example slowly coalescing Galactic Binaries or the inspiral phase of binary black-hole binaries or neutron stars, and is not restricted to the simple case we present in Section~\ref{sec:sol}. For point particles and more general sources, computing the full solutions to higher accuracy than $1$PN is a work in progress \cite{solpaper}, where the solutions will necessarily depend on the location of the source.

\begin{acknowledgments}
This work was financed by CNES and is partly related to LISA. NAN is grateful for discussions with Quentin G. Bailey, Fran\c{c}ois Larrouturou, Adrien Bourgoin, and Samy Aoulad Lafkih. NAN acknowledges support by PSL/Observatoire de Paris.
\end{acknowledgments}

\appendix

\section{Post-Newtonian potentials}\label{app:PNpots}
The Poisson-like equations (\ref{eq:poisson}) have the following general solutions
\begin{equation}\label{eq:potentials}
    \begin{aligned}
        U =& G\int d^3x^\prime\frac{{\rho^*}^\prime}{|\mathbf{x}-\mathbf{x^\prime}|} \\
        \psi =& G\int d^3x^\prime \frac{{\rho^*}^\prime(\tfrac{3}{2}{v^\prime}^2-U^\prime+\Pi^\prime)+3p^\prime}{|\mathbf{x}-\mathbf{x^\prime}|} \\
        V =& G\int d^3x^\prime \frac{{\rho^*}^\prime({v^\prime}^2-\tfrac{1}{2}U^\prime)+3p^\prime}{|\mathbf{x}-\mathbf{x^\prime}|},\\
        X =& G\int d^3x^\prime {\rho^*}^\prime |\mathbf{x}-\mathbf{x^\prime}| \\
        U^j =& G\int d^3x^\prime \frac{{\rho^*}^\prime {v^\prime}^j}{|\mathbf{x}-\mathbf{x^\prime}|} \\
        W^{jk} =& G\int d^3x^\prime \frac{{\rho^*}^\prime ({v^\prime}^j){v^\prime}^k-\tfrac{1}{2}U^\prime\delta^{jk}+p^\prime\delta^{jk}}{|\mathbf{x}-\mathbf{x^\prime}|} \\
        \chi^{jk} =& G^2\int d^3y_1d^3y_2 \frac{\rho^*_1\rho^*_2 (n_1^j-n_{12}^j)(n_2^k+n_{12}^k)}{S^2}\\& -G^2\int d^3y_1d^3y_2\frac{\rho^*_1\rho^*_2(n^j_{12}n^k_{12}-\delta^{jk})}{S r_{12}}.
    \end{aligned}
\end{equation}
Here, the notation reads: $\mathbf{r}_1=|\mathbf{x}-\mathbf{y}_1|$, $\mathbf{n}_1=\mathbf{r}_1/r_1$, $\mathbf{r}_{12}=\mathbf{y}_1-\mathbf{y}_2$, $S=r_1+r_2+r_{12}$.

\vfill

\bibliography{apssamp}

\providecommand{\noopsort}[1]{}\providecommand{\singleletter}[1]{#1}%
\begin{thebibliography}{68}%
\makeatletter
\providecommand \@ifxundefined [1]{%
 \@ifx{#1\undefined}
}%
\providecommand \@ifnum [1]{%
 \ifnum #1\expandafter \@firstoftwo
 \else \expandafter \@secondoftwo
 \fi
}%
\providecommand \@ifx [1]{%
 \ifx #1\expandafter \@firstoftwo
 \else \expandafter \@secondoftwo
 \fi
}%
\providecommand \natexlab [1]{#1}%
\providecommand \enquote  [1]{``#1''}%
\providecommand \bibnamefont  [1]{#1}%
\providecommand \bibfnamefont [1]{#1}%
\providecommand \citenamefont [1]{#1}%
\providecommand \href@noop [0]{\@secondoftwo}%
\providecommand \href [0]{\begingroup \@sanitize@url \@href}%
\providecommand \@href[1]{\@@startlink{#1}\@@href}%
\providecommand \@@href[1]{\endgroup#1\@@endlink}%
\providecommand \@sanitize@url [0]{\catcode `\\12\catcode `\$12\catcode
  `\&12\catcode `\#12\catcode `\^12\catcode `\_12\catcode `\%12\relax}%
\providecommand \@@startlink[1]{}%
\providecommand \@@endlink[0]{}%
\providecommand \url  [0]{\begingroup\@sanitize@url \@url }%
\providecommand \@url [1]{\endgroup\@href {#1}{\urlprefix }}%
\providecommand \urlprefix  [0]{URL }%
\providecommand \Eprint [0]{\href }%
\providecommand \doibase [0]{https://doi.org/}%
\providecommand \selectlanguage [0]{\@gobble}%
\providecommand \bibinfo  [0]{\@secondoftwo}%
\providecommand \bibfield  [0]{\@secondoftwo}%
\providecommand \translation [1]{[#1]}%
\providecommand \BibitemOpen [0]{}%
\providecommand \bibitemStop [0]{}%
\providecommand \bibitemNoStop [0]{.\EOS\space}%
\providecommand \EOS [0]{\spacefactor3000\relax}%
\providecommand \BibitemShut  [1]{\csname bibitem#1\endcsname}%
\let\auto@bib@innerbib\@empty
\bibitem [{\citenamefont {Abbott}\ \emph {et~al.}(2017)\citenamefont {Abbott}
  \emph {et~al.}}]{LIGOScientific:2017zic}%
  \BibitemOpen
  \bibfield  {author} {\bibinfo {author} {\bibfnamefont {B.~P.}\ \bibnamefont
  {Abbott}} \emph {et~al.} (\bibinfo {collaboration} {LIGO Scientific, Virgo,
  Fermi-GBM, INTEGRAL}),\ }\bibfield  {title} {\bibinfo {title} {{Gravitational
  Waves and Gamma-rays from a Binary Neutron Star Merger: GW170817 and GRB
  170817A}},\ }\href {https://doi.org/10.3847/2041-8213/aa920c} {\bibfield
  {journal} {\bibinfo  {journal} {Astrophys. J. Lett.}\ }\textbf {\bibinfo
  {volume} {848}},\ \bibinfo {pages} {L13} (\bibinfo {year} {2017})},\ \Eprint
  {https://arxiv.org/abs/1710.05834} {arXiv:1710.05834 [astro-ph.HE]}
  \BibitemShut {NoStop}%
\bibitem [{\citenamefont {Yagi}\ and\ \citenamefont
  {Yunes}(2016)}]{Yagi:2015pkc}%
  \BibitemOpen
  \bibfield  {author} {\bibinfo {author} {\bibfnamefont {K.}~\bibnamefont
  {Yagi}}\ and\ \bibinfo {author} {\bibfnamefont {N.}~\bibnamefont {Yunes}},\
  }\bibfield  {title} {\bibinfo {title} {{Binary Love Relations}},\ }\href
  {https://doi.org/10.1088/0264-9381/33/13/13LT01} {\bibfield  {journal}
  {\bibinfo  {journal} {Class. Quant. Grav.}\ }\textbf {\bibinfo {volume}
  {33}},\ \bibinfo {pages} {13LT01} (\bibinfo {year} {2016})},\ \Eprint
  {https://arxiv.org/abs/1512.02639} {arXiv:1512.02639 [gr-qc]} \BibitemShut
  {NoStop}%
\bibitem [{\citenamefont {Chatziioannou}\ \emph {et~al.}(2018)\citenamefont
  {Chatziioannou}, \citenamefont {Haster},\ and\ \citenamefont
  {Zimmerman}}]{Chatziioannou:2018vzf}%
  \BibitemOpen
  \bibfield  {author} {\bibinfo {author} {\bibfnamefont {K.}~\bibnamefont
  {Chatziioannou}}, \bibinfo {author} {\bibfnamefont {C.-J.}\ \bibnamefont
  {Haster}},\ and\ \bibinfo {author} {\bibfnamefont {A.}~\bibnamefont
  {Zimmerman}},\ }\bibfield  {title} {\bibinfo {title} {{Measuring the neutron
  star tidal deformability with equation-of-state-independent relations and
  gravitational waves}},\ }\href {https://doi.org/10.1103/PhysRevD.97.104036}
  {\bibfield  {journal} {\bibinfo  {journal} {Phys. Rev. D}\ }\textbf {\bibinfo
  {volume} {97}},\ \bibinfo {pages} {104036} (\bibinfo {year} {2018})},\
  \Eprint {https://arxiv.org/abs/1804.03221} {arXiv:1804.03221 [gr-qc]}
  \BibitemShut {NoStop}%
\bibitem [{\citenamefont {Radice}\ \emph {et~al.}(2018)\citenamefont {Radice},
  \citenamefont {Perego}, \citenamefont {Zappa},\ and\ \citenamefont
  {Bernuzzi}}]{Radice:2017lry}%
  \BibitemOpen
  \bibfield  {author} {\bibinfo {author} {\bibfnamefont {D.}~\bibnamefont
  {Radice}}, \bibinfo {author} {\bibfnamefont {A.}~\bibnamefont {Perego}},
  \bibinfo {author} {\bibfnamefont {F.}~\bibnamefont {Zappa}},\ and\ \bibinfo
  {author} {\bibfnamefont {S.}~\bibnamefont {Bernuzzi}},\ }\bibfield  {title}
  {\bibinfo {title} {{GW170817: Joint Constraint on the Neutron Star Equation
  of State from Multimessenger Observations}},\ }\href
  {https://doi.org/10.3847/2041-8213/aaa402} {\bibfield  {journal} {\bibinfo
  {journal} {Astrophys. J. Lett.}\ }\textbf {\bibinfo {volume} {852}},\
  \bibinfo {pages} {L29} (\bibinfo {year} {2018})},\ \Eprint
  {https://arxiv.org/abs/1711.03647} {arXiv:1711.03647 [astro-ph.HE]}
  \BibitemShut {NoStop}%
\bibitem [{\citenamefont {Kostelecky}\ and\ \citenamefont
  {Samuel}(1989)}]{Kostelecky:1988zi}%
  \BibitemOpen
  \bibfield  {author} {\bibinfo {author} {\bibfnamefont {V.~A.}\ \bibnamefont
  {Kostelecky}}\ and\ \bibinfo {author} {\bibfnamefont {S.}~\bibnamefont
  {Samuel}},\ }\bibfield  {title} {\bibinfo {title} {{Spontaneous Breaking of
  Lorentz Symmetry in String Theory}},\ }\href
  {https://doi.org/10.1103/PhysRevD.39.683} {\bibfield  {journal} {\bibinfo
  {journal} {Phys. Rev. D}\ }\textbf {\bibinfo {volume} {39}},\ \bibinfo
  {pages} {683} (\bibinfo {year} {1989})}\BibitemShut {NoStop}%
\bibitem [{\citenamefont {Kostelecky}\ and\ \citenamefont
  {Potting}(1991)}]{Kostelecky:1991ak}%
  \BibitemOpen
  \bibfield  {author} {\bibinfo {author} {\bibfnamefont {V.~A.}\ \bibnamefont
  {Kostelecky}}\ and\ \bibinfo {author} {\bibfnamefont {R.}~\bibnamefont
  {Potting}},\ }\bibfield  {title} {\bibinfo {title} {{CPT and strings}},\
  }\href {https://doi.org/10.1016/0550-3213(91)90071-5} {\bibfield  {journal}
  {\bibinfo  {journal} {Nucl. Phys. B}\ }\textbf {\bibinfo {volume} {359}},\
  \bibinfo {pages} {545} (\bibinfo {year} {1991})}\BibitemShut {NoStop}%
\bibitem [{\citenamefont {Gambini}\ and\ \citenamefont
  {Pullin}(1999)}]{Gambini:1998it}%
  \BibitemOpen
  \bibfield  {author} {\bibinfo {author} {\bibfnamefont {R.}~\bibnamefont
  {Gambini}}\ and\ \bibinfo {author} {\bibfnamefont {J.}~\bibnamefont
  {Pullin}},\ }\bibfield  {title} {\bibinfo {title} {{Nonstandard optics from
  quantum space-time}},\ }\href {https://doi.org/10.1103/PhysRevD.59.124021}
  {\bibfield  {journal} {\bibinfo  {journal} {Phys. Rev. D}\ }\textbf {\bibinfo
  {volume} {59}},\ \bibinfo {pages} {124021} (\bibinfo {year} {1999})},\
  \Eprint {https://arxiv.org/abs/gr-qc/9809038} {arXiv:gr-qc/9809038}
  \BibitemShut {NoStop}%
\bibitem [{\citenamefont {Carroll}\ \emph {et~al.}(2001)\citenamefont
  {Carroll}, \citenamefont {Harvey}, \citenamefont {Kostelecky}, \citenamefont
  {Lane},\ and\ \citenamefont {Okamoto}}]{Carroll:2001ws}%
  \BibitemOpen
  \bibfield  {author} {\bibinfo {author} {\bibfnamefont {S.~M.}\ \bibnamefont
  {Carroll}}, \bibinfo {author} {\bibfnamefont {J.~A.}\ \bibnamefont {Harvey}},
  \bibinfo {author} {\bibfnamefont {V.~A.}\ \bibnamefont {Kostelecky}},
  \bibinfo {author} {\bibfnamefont {C.~D.}\ \bibnamefont {Lane}},\ and\
  \bibinfo {author} {\bibfnamefont {T.}~\bibnamefont {Okamoto}},\ }\bibfield
  {title} {\bibinfo {title} {{Noncommutative field theory and Lorentz
  violation}},\ }\href {https://doi.org/10.1103/PhysRevLett.87.141601}
  {\bibfield  {journal} {\bibinfo  {journal} {Phys. Rev. Lett.}\ }\textbf
  {\bibinfo {volume} {87}},\ \bibinfo {pages} {141601} (\bibinfo {year}
  {2001})},\ \Eprint {https://arxiv.org/abs/hep-th/0105082}
  {arXiv:hep-th/0105082} \BibitemShut {NoStop}%
\bibitem [{\citenamefont {Addazi}\ \emph {et~al.}(2022)\citenamefont {Addazi}
  \emph {et~al.}}]{Addazi:2021xuf}%
  \BibitemOpen
  \bibfield  {author} {\bibinfo {author} {\bibfnamefont {A.}~\bibnamefont
  {Addazi}} \emph {et~al.},\ }\bibfield  {title} {\bibinfo {title} {{Quantum
  gravity phenomenology at the dawn of the multi-messenger era\textemdash{}A
  review}},\ }\href {https://doi.org/10.1016/j.ppnp.2022.103948} {\bibfield
  {journal} {\bibinfo  {journal} {Prog. Part. Nucl. Phys.}\ }\textbf {\bibinfo
  {volume} {125}},\ \bibinfo {pages} {103948} (\bibinfo {year} {2022})},\
  \Eprint {https://arxiv.org/abs/2111.05659} {arXiv:2111.05659 [hep-ph]}
  \BibitemShut {NoStop}%
\bibitem [{\citenamefont {Mariz}\ \emph {et~al.}(2023)\citenamefont {Mariz},
  \citenamefont {Nascimento},\ and\ \citenamefont {Petrov}}]{Mariz:2022oib}%
  \BibitemOpen
  \bibfield  {author} {\bibinfo {author} {\bibfnamefont {T.}~\bibnamefont
  {Mariz}}, \bibinfo {author} {\bibfnamefont {J.~R.}\ \bibnamefont
  {Nascimento}},\ and\ \bibinfo {author} {\bibfnamefont {A.}~\bibnamefont
  {Petrov}},\ }\href {https://doi.org/10.1007/978-3-031-20120-2} {\emph
  {\bibinfo {title} {{Lorentz Symmetry Breaking\textemdash{}Classical and
  Quantum Aspects}}}},\ SpringerBriefs in Physics\ (\bibinfo  {publisher}
  {Springer},\ \bibinfo {year} {2023})\ \Eprint
  {https://arxiv.org/abs/2205.02594} {arXiv:2205.02594 [hep-th]} \BibitemShut
  {NoStop}%
\bibitem [{\citenamefont {Horava}(2009)}]{Horava:2009uw}%
  \BibitemOpen
  \bibfield  {author} {\bibinfo {author} {\bibfnamefont {P.}~\bibnamefont
  {Horava}},\ }\bibfield  {title} {\bibinfo {title} {{Quantum Gravity at a
  Lifshitz Point}},\ }\href {https://doi.org/10.1103/PhysRevD.79.084008}
  {\bibfield  {journal} {\bibinfo  {journal} {Phys. Rev. D}\ }\textbf {\bibinfo
  {volume} {79}},\ \bibinfo {pages} {084008} (\bibinfo {year} {2009})},\
  \Eprint {https://arxiv.org/abs/0901.3775} {arXiv:0901.3775 [hep-th]}
  \BibitemShut {NoStop}%
\bibitem [{\citenamefont {Abbott}\ \emph {et~al.}(2021)\citenamefont {Abbott}
  \emph {et~al.}}]{LIGOScientific:2021sio}%
  \BibitemOpen
  \bibfield  {author} {\bibinfo {author} {\bibfnamefont {R.}~\bibnamefont
  {Abbott}} \emph {et~al.} (\bibinfo {collaboration} {LIGO Scientific, VIRGO,
  KAGRA}),\ }\bibfield  {title} {\bibinfo {title} {{Tests of General Relativity
  with GWTC-3}},\ }\href@noop {} {\  (\bibinfo {year} {2021})},\ \Eprint
  {https://arxiv.org/abs/2112.06861} {arXiv:2112.06861 [gr-qc]} \BibitemShut
  {NoStop}%
\bibitem [{\citenamefont {Liu}\ \emph {et~al.}(2020)\citenamefont {Liu},
  \citenamefont {He}, \citenamefont {Mikulski}, \citenamefont {Palenova},
  \citenamefont {Williams}, \citenamefont {Creighton},\ and\ \citenamefont
  {Tasson}}]{Liu:2020slm}%
  \BibitemOpen
  \bibfield  {author} {\bibinfo {author} {\bibfnamefont {X.}~\bibnamefont
  {Liu}}, \bibinfo {author} {\bibfnamefont {V.~F.}\ \bibnamefont {He}},
  \bibinfo {author} {\bibfnamefont {T.~M.}\ \bibnamefont {Mikulski}}, \bibinfo
  {author} {\bibfnamefont {D.}~\bibnamefont {Palenova}}, \bibinfo {author}
  {\bibfnamefont {C.~E.}\ \bibnamefont {Williams}}, \bibinfo {author}
  {\bibfnamefont {J.}~\bibnamefont {Creighton}},\ and\ \bibinfo {author}
  {\bibfnamefont {J.~D.}\ \bibnamefont {Tasson}},\ }\bibfield  {title}
  {\bibinfo {title} {{Measuring the speed of gravitational waves from the first
  and second observing run of Advanced LIGO and Advanced Virgo}},\ }\href
  {https://doi.org/10.1103/PhysRevD.102.024028} {\bibfield  {journal} {\bibinfo
   {journal} {Phys. Rev. D}\ }\textbf {\bibinfo {volume} {102}},\ \bibinfo
  {pages} {024028} (\bibinfo {year} {2020})},\ \Eprint
  {https://arxiv.org/abs/2005.03121} {arXiv:2005.03121 [gr-qc]} \BibitemShut
  {NoStop}%
\bibitem [{\citenamefont {Colladay}\ and\ \citenamefont
  {Kostelecky}(1997)}]{Colladay:1996iz}%
  \BibitemOpen
  \bibfield  {author} {\bibinfo {author} {\bibfnamefont {D.}~\bibnamefont
  {Colladay}}\ and\ \bibinfo {author} {\bibfnamefont {V.~A.}\ \bibnamefont
  {Kostelecky}},\ }\bibfield  {title} {\bibinfo {title} {{CPT violation and the
  standard model}},\ }\href {https://doi.org/10.1103/PhysRevD.55.6760}
  {\bibfield  {journal} {\bibinfo  {journal} {Phys. Rev. D}\ }\textbf {\bibinfo
  {volume} {55}},\ \bibinfo {pages} {6760} (\bibinfo {year} {1997})},\ \Eprint
  {https://arxiv.org/abs/hep-ph/9703464} {arXiv:hep-ph/9703464} \BibitemShut
  {NoStop}%
\bibitem [{\citenamefont {Colladay}\ and\ \citenamefont
  {Kostelecky}(1998)}]{Colladay:1998fq}%
  \BibitemOpen
  \bibfield  {author} {\bibinfo {author} {\bibfnamefont {D.}~\bibnamefont
  {Colladay}}\ and\ \bibinfo {author} {\bibfnamefont {V.~A.}\ \bibnamefont
  {Kostelecky}},\ }\bibfield  {title} {\bibinfo {title} {{Lorentz violating
  extension of the standard model}},\ }\href
  {https://doi.org/10.1103/PhysRevD.58.116002} {\bibfield  {journal} {\bibinfo
  {journal} {Phys. Rev. D}\ }\textbf {\bibinfo {volume} {58}},\ \bibinfo
  {pages} {116002} (\bibinfo {year} {1998})},\ \Eprint
  {https://arxiv.org/abs/hep-ph/9809521} {arXiv:hep-ph/9809521} \BibitemShut
  {NoStop}%
\bibitem [{\citenamefont {Kostelecky}(2004)}]{Kostelecky:2003fs}%
  \BibitemOpen
  \bibfield  {author} {\bibinfo {author} {\bibfnamefont {V.~A.}\ \bibnamefont
  {Kostelecky}},\ }\bibfield  {title} {\bibinfo {title} {{Gravity, Lorentz
  violation, and the standard model}},\ }\href
  {https://doi.org/10.1103/PhysRevD.69.105009} {\bibfield  {journal} {\bibinfo
  {journal} {Phys. Rev. D}\ }\textbf {\bibinfo {volume} {69}},\ \bibinfo
  {pages} {105009} (\bibinfo {year} {2004})},\ \Eprint
  {https://arxiv.org/abs/hep-th/0312310} {arXiv:hep-th/0312310} \BibitemShut
  {NoStop}%
\bibitem [{\citenamefont {Bailey}\ and\ \citenamefont
  {Kostelecky}(2006)}]{Bailey:2006fd}%
  \BibitemOpen
  \bibfield  {author} {\bibinfo {author} {\bibfnamefont {Q.~G.}\ \bibnamefont
  {Bailey}}\ and\ \bibinfo {author} {\bibfnamefont {V.~A.}\ \bibnamefont
  {Kostelecky}},\ }\bibfield  {title} {\bibinfo {title} {{Signals for Lorentz
  violation in post-Newtonian gravity}},\ }\href
  {https://doi.org/10.1103/PhysRevD.74.045001} {\bibfield  {journal} {\bibinfo
  {journal} {Phys. Rev. D}\ }\textbf {\bibinfo {volume} {74}},\ \bibinfo
  {pages} {045001} (\bibinfo {year} {2006})},\ \Eprint
  {https://arxiv.org/abs/gr-qc/0603030} {arXiv:gr-qc/0603030} \BibitemShut
  {NoStop}%
\bibitem [{\citenamefont {Kostelecky}\ and\ \citenamefont
  {Tasson}(2011)}]{Kostelecky:2010ze}%
  \BibitemOpen
  \bibfield  {author} {\bibinfo {author} {\bibfnamefont {A.~V.}\ \bibnamefont
  {Kostelecky}}\ and\ \bibinfo {author} {\bibfnamefont {J.~D.}\ \bibnamefont
  {Tasson}},\ }\bibfield  {title} {\bibinfo {title} {{Matter-gravity couplings
  and Lorentz violation}},\ }\href {https://doi.org/10.1103/PhysRevD.83.016013}
  {\bibfield  {journal} {\bibinfo  {journal} {Phys. Rev. D}\ }\textbf {\bibinfo
  {volume} {83}},\ \bibinfo {pages} {016013} (\bibinfo {year} {2011})},\
  \Eprint {https://arxiv.org/abs/1006.4106} {arXiv:1006.4106 [gr-qc]}
  \BibitemShut {NoStop}%
\bibitem [{\citenamefont {Bluhm}(2015)}]{Bluhm:2014oua}%
  \BibitemOpen
  \bibfield  {author} {\bibinfo {author} {\bibfnamefont {R.}~\bibnamefont
  {Bluhm}},\ }\bibfield  {title} {\bibinfo {title} {{Explicit versus
  Spontaneous Diffeomorphism Breaking in Gravity}},\ }\href
  {https://doi.org/10.1103/PhysRevD.91.065034} {\bibfield  {journal} {\bibinfo
  {journal} {Phys. Rev. D}\ }\textbf {\bibinfo {volume} {91}},\ \bibinfo
  {pages} {065034} (\bibinfo {year} {2015})},\ \Eprint
  {https://arxiv.org/abs/1401.4515} {arXiv:1401.4515 [gr-qc]} \BibitemShut
  {NoStop}%
\bibitem [{\citenamefont {Kosteleck\'y}\ and\ \citenamefont
  {Mewes}(2018)}]{Kostelecky:2017zob}%
  \BibitemOpen
  \bibfield  {author} {\bibinfo {author} {\bibfnamefont {V.~A.}\ \bibnamefont
  {Kosteleck\'y}}\ and\ \bibinfo {author} {\bibfnamefont {M.}~\bibnamefont
  {Mewes}},\ }\bibfield  {title} {\bibinfo {title} {{Lorentz and Diffeomorphism
  Violations in Linearized Gravity}},\ }\href
  {https://doi.org/10.1016/j.physletb.2018.01.082} {\bibfield  {journal}
  {\bibinfo  {journal} {Phys. Lett. B}\ }\textbf {\bibinfo {volume} {779}},\
  \bibinfo {pages} {136} (\bibinfo {year} {2018})},\ \Eprint
  {https://arxiv.org/abs/1712.10268} {arXiv:1712.10268 [gr-qc]} \BibitemShut
  {NoStop}%
\bibitem [{\citenamefont {Kosteleck\'y}\ and\ \citenamefont
  {Li}(2021)}]{Kostelecky:2020hbb}%
  \BibitemOpen
  \bibfield  {author} {\bibinfo {author} {\bibfnamefont {V.~A.}\ \bibnamefont
  {Kosteleck\'y}}\ and\ \bibinfo {author} {\bibfnamefont {Z.}~\bibnamefont
  {Li}},\ }\bibfield  {title} {\bibinfo {title} {{Backgrounds in gravitational
  effective field theory}},\ }\href
  {https://doi.org/10.1103/PhysRevD.103.024059} {\bibfield  {journal} {\bibinfo
   {journal} {Phys. Rev. D}\ }\textbf {\bibinfo {volume} {103}},\ \bibinfo
  {pages} {024059} (\bibinfo {year} {2021})},\ \Eprint
  {https://arxiv.org/abs/2008.12206} {arXiv:2008.12206 [gr-qc]} \BibitemShut
  {NoStop}%
\bibitem [{\citenamefont {O'Neal-Ault}\ \emph
  {et~al.}(2021{\natexlab{a}})\citenamefont {O'Neal-Ault}, \citenamefont
  {Bailey},\ and\ \citenamefont {Nilsson}}]{ONeal-Ault:2020ebv}%
  \BibitemOpen
  \bibfield  {author} {\bibinfo {author} {\bibfnamefont {K.}~\bibnamefont
  {O'Neal-Ault}}, \bibinfo {author} {\bibfnamefont {Q.~G.}\ \bibnamefont
  {Bailey}},\ and\ \bibinfo {author} {\bibfnamefont {N.~A.}\ \bibnamefont
  {Nilsson}},\ }\bibfield  {title} {\bibinfo {title} {{3+1 formulation of the
  standard model extension gravity sector}},\ }\href
  {https://doi.org/10.1103/PhysRevD.103.044010} {\bibfield  {journal} {\bibinfo
   {journal} {Phys. Rev. D}\ }\textbf {\bibinfo {volume} {103}},\ \bibinfo
  {pages} {044010} (\bibinfo {year} {2021}{\natexlab{a}})},\ \Eprint
  {https://arxiv.org/abs/2009.00949} {arXiv:2009.00949 [gr-qc]} \BibitemShut
  {NoStop}%
\bibitem [{\citenamefont {Haegel}\ \emph {et~al.}(2023)\citenamefont {Haegel},
  \citenamefont {O'Neal-Ault}, \citenamefont {Bailey}, \citenamefont {Tasson},
  \citenamefont {Bloom},\ and\ \citenamefont {Shao}}]{Haegel:2022ymk}%
  \BibitemOpen
  \bibfield  {author} {\bibinfo {author} {\bibfnamefont {L.}~\bibnamefont
  {Haegel}}, \bibinfo {author} {\bibfnamefont {K.}~\bibnamefont {O'Neal-Ault}},
  \bibinfo {author} {\bibfnamefont {Q.~G.}\ \bibnamefont {Bailey}}, \bibinfo
  {author} {\bibfnamefont {J.~D.}\ \bibnamefont {Tasson}}, \bibinfo {author}
  {\bibfnamefont {M.}~\bibnamefont {Bloom}},\ and\ \bibinfo {author}
  {\bibfnamefont {L.}~\bibnamefont {Shao}},\ }\bibfield  {title} {\bibinfo
  {title} {{Search for anisotropic, birefringent spacetime-symmetry breaking in
  gravitational wave propagation from GWTC-3}},\ }\href
  {https://doi.org/10.1103/PhysRevD.107.064031} {\bibfield  {journal} {\bibinfo
   {journal} {Phys. Rev. D}\ }\textbf {\bibinfo {volume} {107}},\ \bibinfo
  {pages} {064031} (\bibinfo {year} {2023})},\ \Eprint
  {https://arxiv.org/abs/2210.04481} {arXiv:2210.04481 [gr-qc]} \BibitemShut
  {NoStop}%
\bibitem [{\citenamefont {Kosteleck\'y}\ and\ \citenamefont
  {Mewes}(2016)}]{Kostelecky:2016kfm}%
  \BibitemOpen
  \bibfield  {author} {\bibinfo {author} {\bibfnamefont {V.~A.}\ \bibnamefont
  {Kosteleck\'y}}\ and\ \bibinfo {author} {\bibfnamefont {M.}~\bibnamefont
  {Mewes}},\ }\bibfield  {title} {\bibinfo {title} {{Testing local Lorentz
  invariance with gravitational waves}},\ }\href
  {https://doi.org/10.1016/j.physletb.2016.04.040} {\bibfield  {journal}
  {\bibinfo  {journal} {Phys. Lett. B}\ }\textbf {\bibinfo {volume} {757}},\
  \bibinfo {pages} {510} (\bibinfo {year} {2016})},\ \Eprint
  {https://arxiv.org/abs/1602.04782} {arXiv:1602.04782 [gr-qc]} \BibitemShut
  {NoStop}%
\bibitem [{\citenamefont {Kosteleck\'y}\ and\ \citenamefont
  {Tasson}(2015)}]{Kostelecky:2015dpa}%
  \BibitemOpen
  \bibfield  {author} {\bibinfo {author} {\bibfnamefont {V.~A.}\ \bibnamefont
  {Kosteleck\'y}}\ and\ \bibinfo {author} {\bibfnamefont {J.~D.}\ \bibnamefont
  {Tasson}},\ }\bibfield  {title} {\bibinfo {title} {{Constraints on Lorentz
  violation from gravitational \v{C}erenkov radiation}},\ }\href
  {https://doi.org/10.1016/j.physletb.2015.08.060} {\bibfield  {journal}
  {\bibinfo  {journal} {Phys. Lett. B}\ }\textbf {\bibinfo {volume} {749}},\
  \bibinfo {pages} {551} (\bibinfo {year} {2015})},\ \Eprint
  {https://arxiv.org/abs/1508.07007} {arXiv:1508.07007 [gr-qc]} \BibitemShut
  {NoStop}%
\bibitem [{\citenamefont {Wang}\ \emph {et~al.}(2021)\citenamefont {Wang},
  \citenamefont {Shao},\ and\ \citenamefont {Liu}}]{Wang:2021ctl}%
  \BibitemOpen
  \bibfield  {author} {\bibinfo {author} {\bibfnamefont {Z.}~\bibnamefont
  {Wang}}, \bibinfo {author} {\bibfnamefont {L.}~\bibnamefont {Shao}},\ and\
  \bibinfo {author} {\bibfnamefont {C.}~\bibnamefont {Liu}},\ }\bibfield
  {title} {\bibinfo {title} {{New Limits on the Lorentz/CPT Symmetry Through 50
  Gravitational-wave Events}},\ }\href
  {https://doi.org/10.3847/1538-4357/ac223c} {\bibfield  {journal} {\bibinfo
  {journal} {Astrophys. J.}\ }\textbf {\bibinfo {volume} {921}},\ \bibinfo
  {pages} {158} (\bibinfo {year} {2021})},\ \Eprint
  {https://arxiv.org/abs/2108.02974} {arXiv:2108.02974 [gr-qc]} \BibitemShut
  {NoStop}%
\bibitem [{\citenamefont {Niu}\ \emph {et~al.}(2022)\citenamefont {Niu},
  \citenamefont {Zhu},\ and\ \citenamefont {Zhao}}]{Niu:2022yhr}%
  \BibitemOpen
  \bibfield  {author} {\bibinfo {author} {\bibfnamefont {R.}~\bibnamefont
  {Niu}}, \bibinfo {author} {\bibfnamefont {T.}~\bibnamefont {Zhu}},\ and\
  \bibinfo {author} {\bibfnamefont {W.}~\bibnamefont {Zhao}},\ }\bibfield
  {title} {\bibinfo {title} {{Testing Lorentz invariance of gravity in the
  Standard-Model Extension with GWTC-3}},\ }\href
  {https://doi.org/10.1088/1475-7516/2022/12/011} {\bibfield  {journal}
  {\bibinfo  {journal} {JCAP}\ }\textbf {\bibinfo {volume} {12}},\ \bibinfo
  {pages} {011}},\ \Eprint {https://arxiv.org/abs/2202.05092} {arXiv:2202.05092
  [gr-qc]} \BibitemShut {NoStop}%
\bibitem [{\citenamefont {Nilsson}(2022)}]{Nilsson:2022mzq}%
  \BibitemOpen
  \bibfield  {author} {\bibinfo {author} {\bibfnamefont {N.~A.}\ \bibnamefont
  {Nilsson}},\ }\bibfield  {title} {\bibinfo {title} {{Explicit
  spacetime-symmetry breaking and the dynamics of primordial fields}},\ }\href
  {https://doi.org/10.1103/PhysRevD.106.104036} {\bibfield  {journal} {\bibinfo
   {journal} {Phys. Rev. D}\ }\textbf {\bibinfo {volume} {106}},\ \bibinfo
  {pages} {104036} (\bibinfo {year} {2022})},\ \Eprint
  {https://arxiv.org/abs/2205.00496} {arXiv:2205.00496 [gr-qc]} \BibitemShut
  {NoStop}%
\bibitem [{\citenamefont {Iorio}(2012)}]{Iorio:2012gr}%
  \BibitemOpen
  \bibfield  {author} {\bibinfo {author} {\bibfnamefont {L.}~\bibnamefont
  {Iorio}},\ }\bibfield  {title} {\bibinfo {title} {{Orbital effects of
  Lorentz-violating Standard Model Extension gravitomagnetism around a static
  body: a sensitivity analysis}},\ }\href
  {https://doi.org/10.1088/0264-9381/29/17/175007} {\bibfield  {journal}
  {\bibinfo  {journal} {Class. Quant. Grav.}\ }\textbf {\bibinfo {volume}
  {29}},\ \bibinfo {pages} {175007} (\bibinfo {year} {2012})},\ \Eprint
  {https://arxiv.org/abs/1203.1859} {arXiv:1203.1859 [gr-qc]} \BibitemShut
  {NoStop}%
\bibitem [{\citenamefont {Hees}\ \emph {et~al.}(2015)\citenamefont {Hees},
  \citenamefont {Bailey}, \citenamefont {Le~Poncin-Lafitte}, \citenamefont
  {Bourgoin}, \citenamefont {Rivoldini}, \citenamefont {Lamine}, \citenamefont
  {Meynadier}, \citenamefont {Guerlin},\ and\ \citenamefont
  {Wolf}}]{Hees:2015mga}%
  \BibitemOpen
  \bibfield  {author} {\bibinfo {author} {\bibfnamefont {A.}~\bibnamefont
  {Hees}}, \bibinfo {author} {\bibfnamefont {Q.~G.}\ \bibnamefont {Bailey}},
  \bibinfo {author} {\bibfnamefont {C.}~\bibnamefont {Le~Poncin-Lafitte}},
  \bibinfo {author} {\bibfnamefont {A.}~\bibnamefont {Bourgoin}}, \bibinfo
  {author} {\bibfnamefont {A.}~\bibnamefont {Rivoldini}}, \bibinfo {author}
  {\bibfnamefont {B.}~\bibnamefont {Lamine}}, \bibinfo {author} {\bibfnamefont
  {F.}~\bibnamefont {Meynadier}}, \bibinfo {author} {\bibfnamefont
  {C.}~\bibnamefont {Guerlin}},\ and\ \bibinfo {author} {\bibfnamefont
  {P.}~\bibnamefont {Wolf}},\ }\bibfield  {title} {\bibinfo {title} {{Testing
  Lorentz symmetry with planetary orbital dynamics}},\ }\href
  {https://doi.org/10.1103/PhysRevD.92.064049} {\bibfield  {journal} {\bibinfo
  {journal} {Phys. Rev. D}\ }\textbf {\bibinfo {volume} {92}},\ \bibinfo
  {pages} {064049} (\bibinfo {year} {2015})},\ \Eprint
  {https://arxiv.org/abs/1508.03478} {arXiv:1508.03478 [gr-qc]} \BibitemShut
  {NoStop}%
\bibitem [{\citenamefont {Le~Poncin-Lafitte}\ \emph {et~al.}(2016)\citenamefont
  {Le~Poncin-Lafitte}, \citenamefont {Hees},\ and\ \citenamefont
  {Lambert}}]{LePoncin-Lafitte:2016ocy}%
  \BibitemOpen
  \bibfield  {author} {\bibinfo {author} {\bibfnamefont {C.}~\bibnamefont
  {Le~Poncin-Lafitte}}, \bibinfo {author} {\bibfnamefont {A.}~\bibnamefont
  {Hees}},\ and\ \bibinfo {author} {\bibfnamefont {S.}~\bibnamefont
  {Lambert}},\ }\bibfield  {title} {\bibinfo {title} {{Lorentz symmetry and
  Very Long Baseline Interferometry}},\ }\href
  {https://doi.org/10.1103/PhysRevD.94.125030} {\bibfield  {journal} {\bibinfo
  {journal} {Phys. Rev. D}\ }\textbf {\bibinfo {volume} {94}},\ \bibinfo
  {pages} {125030} (\bibinfo {year} {2016})},\ \Eprint
  {https://arxiv.org/abs/1604.01663} {arXiv:1604.01663 [gr-qc]} \BibitemShut
  {NoStop}%
\bibitem [{\citenamefont {Bourgoin}\ \emph {et~al.}(2016)\citenamefont
  {Bourgoin}, \citenamefont {Hees}, \citenamefont {Bouquillon}, \citenamefont
  {Le~Poncin-Lafitte}, \citenamefont {Francou},\ and\ \citenamefont
  {Angonin}}]{Bourgoin:2016ynf}%
  \BibitemOpen
  \bibfield  {author} {\bibinfo {author} {\bibfnamefont {A.}~\bibnamefont
  {Bourgoin}}, \bibinfo {author} {\bibfnamefont {A.}~\bibnamefont {Hees}},
  \bibinfo {author} {\bibfnamefont {S.}~\bibnamefont {Bouquillon}}, \bibinfo
  {author} {\bibfnamefont {C.}~\bibnamefont {Le~Poncin-Lafitte}}, \bibinfo
  {author} {\bibfnamefont {G.}~\bibnamefont {Francou}},\ and\ \bibinfo {author}
  {\bibfnamefont {M.~C.}\ \bibnamefont {Angonin}},\ }\bibfield  {title}
  {\bibinfo {title} {{Testing Lorentz symmetry with Lunar Laser Ranging}},\
  }\href {https://doi.org/10.1103/PhysRevLett.117.241301} {\bibfield  {journal}
  {\bibinfo  {journal} {Phys. Rev. Lett.}\ }\textbf {\bibinfo {volume} {117}},\
  \bibinfo {pages} {241301} (\bibinfo {year} {2016})},\ \Eprint
  {https://arxiv.org/abs/1607.00294} {arXiv:1607.00294 [gr-qc]} \BibitemShut
  {NoStop}%
\bibitem [{\citenamefont {Bourgoin}\ \emph {et~al.}(2017)\citenamefont
  {Bourgoin}, \citenamefont {Le~Poncin-Lafitte}, \citenamefont {Hees},
  \citenamefont {Bouquillon}, \citenamefont {Francou},\ and\ \citenamefont
  {Angonin}}]{Bourgoin:2017fpo}%
  \BibitemOpen
  \bibfield  {author} {\bibinfo {author} {\bibfnamefont {A.}~\bibnamefont
  {Bourgoin}}, \bibinfo {author} {\bibfnamefont {C.}~\bibnamefont
  {Le~Poncin-Lafitte}}, \bibinfo {author} {\bibfnamefont {A.}~\bibnamefont
  {Hees}}, \bibinfo {author} {\bibfnamefont {S.}~\bibnamefont {Bouquillon}},
  \bibinfo {author} {\bibfnamefont {G.}~\bibnamefont {Francou}},\ and\ \bibinfo
  {author} {\bibfnamefont {M.-C.}\ \bibnamefont {Angonin}},\ }\bibfield
  {title} {\bibinfo {title} {{Lorentz Symmetry Violations from Matter-Gravity
  Couplings with Lunar Laser Ranging}},\ }\href
  {https://doi.org/10.1103/PhysRevLett.119.201102} {\bibfield  {journal}
  {\bibinfo  {journal} {Phys. Rev. Lett.}\ }\textbf {\bibinfo {volume} {119}},\
  \bibinfo {pages} {201102} (\bibinfo {year} {2017})},\ \Eprint
  {https://arxiv.org/abs/1706.06294} {arXiv:1706.06294 [gr-qc]} \BibitemShut
  {NoStop}%
\bibitem [{\citenamefont {Bourgoin}\ \emph {et~al.}(2021)\citenamefont
  {Bourgoin} \emph {et~al.}}]{Bourgoin:2020ckq}%
  \BibitemOpen
  \bibfield  {author} {\bibinfo {author} {\bibfnamefont {A.}~\bibnamefont
  {Bourgoin}} \emph {et~al.},\ }\bibfield  {title} {\bibinfo {title}
  {{Constraining velocity-dependent Lorentz and $CPT$ violations using lunar
  laser ranging}},\ }\href {https://doi.org/10.1103/PhysRevD.103.064055}
  {\bibfield  {journal} {\bibinfo  {journal} {Phys. Rev. D}\ }\textbf {\bibinfo
  {volume} {103}},\ \bibinfo {pages} {064055} (\bibinfo {year} {2021})},\
  \Eprint {https://arxiv.org/abs/2011.06641} {arXiv:2011.06641 [gr-qc]}
  \BibitemShut {NoStop}%
\bibitem [{\citenamefont {Shao}(2014)}]{Shao:2014oha}%
  \BibitemOpen
  \bibfield  {author} {\bibinfo {author} {\bibfnamefont {L.}~\bibnamefont
  {Shao}},\ }\bibfield  {title} {\bibinfo {title} {{Tests of local Lorentz
  invariance violation of gravity in the standard model extension with
  pulsars}},\ }\href {https://doi.org/10.1103/PhysRevLett.112.111103}
  {\bibfield  {journal} {\bibinfo  {journal} {Phys. Rev. Lett.}\ }\textbf
  {\bibinfo {volume} {112}},\ \bibinfo {pages} {111103} (\bibinfo {year}
  {2014})},\ \Eprint {https://arxiv.org/abs/1402.6452} {arXiv:1402.6452
  [gr-qc]} \BibitemShut {NoStop}%
\bibitem [{\citenamefont {Shao}\ and\ \citenamefont
  {Bailey}(2018)}]{Shao:2018vul}%
  \BibitemOpen
  \bibfield  {author} {\bibinfo {author} {\bibfnamefont {L.}~\bibnamefont
  {Shao}}\ and\ \bibinfo {author} {\bibfnamefont {Q.~G.}\ \bibnamefont
  {Bailey}},\ }\bibfield  {title} {\bibinfo {title} {{Testing
  velocity-dependent CPT-violating gravitational forces with radio pulsars}},\
  }\href {https://doi.org/10.1103/PhysRevD.98.084049} {\bibfield  {journal}
  {\bibinfo  {journal} {Phys. Rev. D}\ }\textbf {\bibinfo {volume} {98}},\
  \bibinfo {pages} {084049} (\bibinfo {year} {2018})},\ \Eprint
  {https://arxiv.org/abs/1810.06332} {arXiv:1810.06332 [gr-qc]} \BibitemShut
  {NoStop}%
\bibitem [{\citenamefont {Kostelecky}\ and\ \citenamefont
  {Russell}(2011)}]{Kostelecky:2008ts}%
  \BibitemOpen
  \bibfield  {author} {\bibinfo {author} {\bibfnamefont {V.~A.}\ \bibnamefont
  {Kostelecky}}\ and\ \bibinfo {author} {\bibfnamefont {N.}~\bibnamefont
  {Russell}},\ }\bibfield  {title} {\bibinfo {title} {{Data Tables for Lorentz
  and CPT Violation}},\ }\href {https://doi.org/10.1103/RevModPhys.83.11}
  {\bibfield  {journal} {\bibinfo  {journal} {Rev. Mod. Phys.}\ }\textbf
  {\bibinfo {volume} {83}},\ \bibinfo {pages} {11} (\bibinfo {year} {2011})},\
  \Eprint {https://arxiv.org/abs/0801.0287} {arXiv:0801.0287 [hep-ph]}
  \BibitemShut {NoStop}%
\bibitem [{\citenamefont {Bailey}\ and\ \citenamefont
  {Lane}(2018)}]{Bailey:2018ifc}%
  \BibitemOpen
  \bibfield  {author} {\bibinfo {author} {\bibfnamefont {Q.~G.}\ \bibnamefont
  {Bailey}}\ and\ \bibinfo {author} {\bibfnamefont {C.~D.}\ \bibnamefont
  {Lane}},\ }\bibfield  {title} {\bibinfo {title} {{Relating Noncommutative
  SO(2,3)* Gravity to the Lorentz-Violating Standard-Model Extension}},\ }\href
  {https://doi.org/10.3390/sym10100480} {\bibfield  {journal} {\bibinfo
  {journal} {Symmetry}\ }\textbf {\bibinfo {volume} {10}},\ \bibinfo {pages}
  {480} (\bibinfo {year} {2018})},\ \Eprint {https://arxiv.org/abs/1810.05136}
  {arXiv:1810.05136 [hep-th]} \BibitemShut {NoStop}%
\bibitem [{\citenamefont {Bluhm}(2010)}]{Bluhm:2010mi}%
  \BibitemOpen
  \bibfield  {author} {\bibinfo {author} {\bibfnamefont {R.}~\bibnamefont
  {Bluhm}},\ }\bibfield  {title} {\bibinfo {title} {{Spontaneous Lorentz
  Violation, Nambu-Goldstone Modes, and Massive Modes}},\ }in\ \href
  {https://doi.org/10.1142/9789814327688_0025} {\emph {\bibinfo {booktitle}
  {{5th Meeting on CPT and Lorentz Symmetry}}}}\ (\bibinfo {year} {2010})\ pp.\
  \bibinfo {pages} {128--132},\ \Eprint {https://arxiv.org/abs/1008.1805}
  {arXiv:1008.1805 [hep-th]} \BibitemShut {NoStop}%
\bibitem [{\citenamefont {Khodadi}\ and\ \citenamefont
  {Schreck}(2023)}]{Khodadi:2023ezj}%
  \BibitemOpen
  \bibfield  {author} {\bibinfo {author} {\bibfnamefont {M.}~\bibnamefont
  {Khodadi}}\ and\ \bibinfo {author} {\bibfnamefont {M.}~\bibnamefont
  {Schreck}},\ }\bibfield  {title} {\bibinfo {title} {{Hubble tension as a
  guide for refining the early Universe: Cosmologies with explicit local
  Lorentz and diffeomorphism violation}},\ }\href
  {https://doi.org/10.1016/j.dark.2023.101170} {\bibfield  {journal} {\bibinfo
  {journal} {Phys. Dark Univ.}\ }\textbf {\bibinfo {volume} {39}},\ \bibinfo
  {pages} {101170} (\bibinfo {year} {2023})},\ \Eprint
  {https://arxiv.org/abs/2301.03883} {arXiv:2301.03883 [gr-qc]} \BibitemShut
  {NoStop}%
\bibitem [{\citenamefont {Reyes}\ and\ \citenamefont
  {Schreck}(2022)}]{Reyes:2022mvm}%
  \BibitemOpen
  \bibfield  {author} {\bibinfo {author} {\bibfnamefont {C.~M.}\ \bibnamefont
  {Reyes}}\ and\ \bibinfo {author} {\bibfnamefont {M.}~\bibnamefont
  {Schreck}},\ }\bibfield  {title} {\bibinfo {title} {{Modified-gravity
  theories with nondynamical background fields}},\ }\href
  {https://doi.org/10.1103/PhysRevD.106.044050} {\bibfield  {journal} {\bibinfo
   {journal} {Phys. Rev. D}\ }\textbf {\bibinfo {volume} {106}},\ \bibinfo
  {pages} {044050} (\bibinfo {year} {2022})},\ \Eprint
  {https://arxiv.org/abs/2202.11881} {arXiv:2202.11881 [hep-th]} \BibitemShut
  {NoStop}%
\bibitem [{\citenamefont {Froggatt}\ and\ \citenamefont
  {Nielsen}(2002)}]{Froggatt:2002gf}%
  \BibitemOpen
  \bibfield  {author} {\bibinfo {author} {\bibfnamefont {C.~D.}\ \bibnamefont
  {Froggatt}}\ and\ \bibinfo {author} {\bibfnamefont {H.~B.}\ \bibnamefont
  {Nielsen}},\ }\bibfield  {title} {\bibinfo {title} {{Derivation of Lorentz
  invariance and three space dimensions in generic field theory}},\ }\href@noop
  {} {\bibfield  {journal} {\bibinfo  {journal} {Bled Workshops Phys}\ }\textbf
  {\bibinfo {volume} {3}},\ \bibinfo {pages} {1} (\bibinfo {year} {2002})},\
  \Eprint {https://arxiv.org/abs/hep-ph/0211106} {arXiv:hep-ph/0211106}
  \BibitemShut {NoStop}%
\bibitem [{\citenamefont {Mavromatos}(2005)}]{Mavromatos:2004sz}%
  \BibitemOpen
  \bibfield  {author} {\bibinfo {author} {\bibfnamefont {N.~E.}\ \bibnamefont
  {Mavromatos}},\ }\bibfield  {title} {\bibinfo {title} {{CPT violation and
  decoherence in quantum gravity}},\ }\href
  {https://doi.org/10.1007/11377306_8} {\bibfield  {journal} {\bibinfo
  {journal} {Lect. Notes Phys.}\ }\textbf {\bibinfo {volume} {669}},\ \bibinfo
  {pages} {245} (\bibinfo {year} {2005})},\ \Eprint
  {https://arxiv.org/abs/gr-qc/0407005} {arXiv:gr-qc/0407005} \BibitemShut
  {NoStop}%
\bibitem [{\citenamefont {Amelino-Camelia}(2002)}]{Amelino-Camelia:2002aqz}%
  \BibitemOpen
  \bibfield  {author} {\bibinfo {author} {\bibfnamefont {G.}~\bibnamefont
  {Amelino-Camelia}},\ }\bibfield  {title} {\bibinfo {title} {{Quantum gravity
  phenomenology: Status and prospects}},\ }\href
  {https://doi.org/10.1142/S0217732302007612} {\bibfield  {journal} {\bibinfo
  {journal} {Mod. Phys. Lett. A}\ }\textbf {\bibinfo {volume} {17}},\ \bibinfo
  {pages} {899} (\bibinfo {year} {2002})},\ \Eprint
  {https://arxiv.org/abs/gr-qc/0204051} {arXiv:gr-qc/0204051} \BibitemShut
  {NoStop}%
\bibitem [{\citenamefont {Barausse}\ \emph {et~al.}(2020)\citenamefont
  {Barausse} \emph {et~al.}}]{Barausse:2020rsu}%
  \BibitemOpen
  \bibfield  {author} {\bibinfo {author} {\bibfnamefont {E.}~\bibnamefont
  {Barausse}} \emph {et~al.},\ }\bibfield  {title} {\bibinfo {title}
  {{Prospects for Fundamental Physics with LISA}},\ }\href
  {https://doi.org/10.1007/s10714-020-02691-1} {\bibfield  {journal} {\bibinfo
  {journal} {Gen. Rel. Grav.}\ }\textbf {\bibinfo {volume} {52}},\ \bibinfo
  {pages} {81} (\bibinfo {year} {2020})},\ \Eprint
  {https://arxiv.org/abs/2001.09793} {arXiv:2001.09793 [gr-qc]} \BibitemShut
  {NoStop}%
\bibitem [{\citenamefont {Arun}\ \emph {et~al.}(2022)\citenamefont {Arun} \emph
  {et~al.}}]{LISA:2022kgy}%
  \BibitemOpen
  \bibfield  {author} {\bibinfo {author} {\bibfnamefont {K.~G.}\ \bibnamefont
  {Arun}} \emph {et~al.} (\bibinfo {collaboration} {LISA}),\ }\bibfield
  {title} {\bibinfo {title} {{New horizons for fundamental physics with
  LISA}},\ }\href {https://doi.org/10.1007/s41114-022-00036-9} {\bibfield
  {journal} {\bibinfo  {journal} {Living Rev. Rel.}\ }\textbf {\bibinfo
  {volume} {25}},\ \bibinfo {pages} {4} (\bibinfo {year} {2022})},\ \Eprint
  {https://arxiv.org/abs/2205.01597} {arXiv:2205.01597 [gr-qc]} \BibitemShut
  {NoStop}%
\bibitem [{\citenamefont {Cornish}\ and\ \citenamefont
  {Robson}(2017)}]{Cornish:2017vip}%
  \BibitemOpen
  \bibfield  {author} {\bibinfo {author} {\bibfnamefont {N.}~\bibnamefont
  {Cornish}}\ and\ \bibinfo {author} {\bibfnamefont {T.}~\bibnamefont
  {Robson}},\ }\bibfield  {title} {\bibinfo {title} {{Galactic binary science
  with the new LISA design}},\ }\href
  {https://doi.org/10.1088/1742-6596/840/1/012024} {\bibfield  {journal}
  {\bibinfo  {journal} {J. Phys. Conf. Ser.}\ }\textbf {\bibinfo {volume}
  {840}},\ \bibinfo {pages} {012024} (\bibinfo {year} {2017})},\ \Eprint
  {https://arxiv.org/abs/1703.09858} {arXiv:1703.09858 [astro-ph.IM]}
  \BibitemShut {NoStop}%
\bibitem [{\citenamefont {Amarilo}\ \emph {et~al.}(2019)\citenamefont
  {Amarilo}, \citenamefont {Barroso}, \citenamefont {Filho},\ and\
  \citenamefont {Maluf}}]{Amarilo:2019lfq}%
  \BibitemOpen
  \bibfield  {author} {\bibinfo {author} {\bibfnamefont {K.~M.}\ \bibnamefont
  {Amarilo}}, \bibinfo {author} {\bibfnamefont {M.}~\bibnamefont {Barroso}},
  \bibinfo {author} {\bibfnamefont {F.}~\bibnamefont {Filho}},\ and\ \bibinfo
  {author} {\bibfnamefont {R.~V.}\ \bibnamefont {Maluf}},\ }\bibfield  {title}
  {\bibinfo {title} {{Modification in Gravitational Waves Production Triggered
  by Spontaneous Lorentz Violation}},\ }\href
  {https://doi.org/10.22323/1.329.0015} {\bibfield  {journal} {\bibinfo
  {journal} {PoS}\ }\textbf {\bibinfo {volume} {BHCB2018}},\ \bibinfo {pages}
  {015} (\bibinfo {year} {2019})}\BibitemShut {NoStop}%
\bibitem [{\citenamefont {Amarilo}\ \emph {et~al.}(2023)\citenamefont
  {Amarilo}, \citenamefont {Filho}, \citenamefont {Filho},\ and\ \citenamefont
  {Reis}}]{Amarilo:2023wpn}%
  \BibitemOpen
  \bibfield  {author} {\bibinfo {author} {\bibfnamefont {K.~M.}\ \bibnamefont
  {Amarilo}}, \bibinfo {author} {\bibfnamefont {M.~B.~F.}\ \bibnamefont
  {Filho}}, \bibinfo {author} {\bibfnamefont {A.~A.~A.}\ \bibnamefont
  {Filho}},\ and\ \bibinfo {author} {\bibfnamefont {J.~A. A.~S.}\ \bibnamefont
  {Reis}},\ }\bibfield  {title} {\bibinfo {title} {{Gravitational waves effects
  in a Lorentz-violating scenario}},\ }\href@noop {} {\  (\bibinfo {year}
  {2023})},\ \Eprint {https://arxiv.org/abs/2307.10937} {arXiv:2307.10937
  [gr-qc]} \BibitemShut {NoStop}%
\bibitem [{\citenamefont {Alexander}\ and\ \citenamefont
  {Yunes}(2009)}]{Alexander:2009tp}%
  \BibitemOpen
  \bibfield  {author} {\bibinfo {author} {\bibfnamefont {S.}~\bibnamefont
  {Alexander}}\ and\ \bibinfo {author} {\bibfnamefont {N.}~\bibnamefont
  {Yunes}},\ }\bibfield  {title} {\bibinfo {title} {{Chern-Simons Modified
  General Relativity}},\ }\href {https://doi.org/10.1016/j.physrep.2009.07.002}
  {\bibfield  {journal} {\bibinfo  {journal} {Phys. Rept.}\ }\textbf {\bibinfo
  {volume} {480}},\ \bibinfo {pages} {1} (\bibinfo {year} {2009})},\ \Eprint
  {https://arxiv.org/abs/0907.2562} {arXiv:0907.2562 [hep-th]} \BibitemShut
  {NoStop}%
\bibitem [{\citenamefont {Poisson}\ and\ \citenamefont
  {Will}(2014)}]{poisson_will_2014}%
  \BibitemOpen
  \bibfield  {author} {\bibinfo {author} {\bibfnamefont {E.}~\bibnamefont
  {Poisson}}\ and\ \bibinfo {author} {\bibfnamefont {C.~M.}\ \bibnamefont
  {Will}},\ }\href {https://doi.org/10.1017/CBO9781139507486} {\emph {\bibinfo
  {title} {Gravity: Newtonian, Post-Newtonian, Relativistic}}}\ (\bibinfo
  {publisher} {Cambridge University Press},\ \bibinfo {year}
  {2014})\BibitemShut {NoStop}%
\bibitem [{\citenamefont {Mewes}(2019)}]{Mewes:2019dhj}%
  \BibitemOpen
  \bibfield  {author} {\bibinfo {author} {\bibfnamefont {M.}~\bibnamefont
  {Mewes}},\ }\bibfield  {title} {\bibinfo {title} {{Signals for Lorentz
  violation in gravitational waves}},\ }\href
  {https://doi.org/10.1103/PhysRevD.99.104062} {\bibfield  {journal} {\bibinfo
  {journal} {Phys. Rev. D}\ }\textbf {\bibinfo {volume} {99}},\ \bibinfo
  {pages} {104062} (\bibinfo {year} {2019})},\ \Eprint
  {https://arxiv.org/abs/1905.00409} {arXiv:1905.00409 [gr-qc]} \BibitemShut
  {NoStop}%
\bibitem [{\citenamefont {O'Neal-Ault}\ \emph
  {et~al.}(2021{\natexlab{b}})\citenamefont {O'Neal-Ault}, \citenamefont
  {Bailey}, \citenamefont {Dumerchat}, \citenamefont {Haegel},\ and\
  \citenamefont {Tasson}}]{ONeal-Ault:2021uwu}%
  \BibitemOpen
  \bibfield  {author} {\bibinfo {author} {\bibfnamefont {K.}~\bibnamefont
  {O'Neal-Ault}}, \bibinfo {author} {\bibfnamefont {Q.~G.}\ \bibnamefont
  {Bailey}}, \bibinfo {author} {\bibfnamefont {T.}~\bibnamefont {Dumerchat}},
  \bibinfo {author} {\bibfnamefont {L.}~\bibnamefont {Haegel}},\ and\ \bibinfo
  {author} {\bibfnamefont {J.}~\bibnamefont {Tasson}},\ }\bibfield  {title}
  {\bibinfo {title} {{Analysis of Birefringence and Dispersion Effects from
  Spacetime-Symmetry Breaking in Gravitational Waves}},\ }\href
  {https://doi.org/10.3390/universe7100380} {\bibfield  {journal} {\bibinfo
  {journal} {Universe}\ }\textbf {\bibinfo {volume} {7}},\ \bibinfo {pages}
  {380} (\bibinfo {year} {2021}{\natexlab{b}})},\ \Eprint
  {https://arxiv.org/abs/2108.06298} {arXiv:2108.06298 [gr-qc]} \BibitemShut
  {NoStop}%
\bibitem [{\citenamefont {Amelino-Camelia}\ \emph {et~al.}(1998)\citenamefont
  {Amelino-Camelia}, \citenamefont {Ellis}, \citenamefont {Mavromatos},
  \citenamefont {Nanopoulos},\ and\ \citenamefont
  {Sarkar}}]{Amelino-Camelia:1997ieq}%
  \BibitemOpen
  \bibfield  {author} {\bibinfo {author} {\bibfnamefont {G.}~\bibnamefont
  {Amelino-Camelia}}, \bibinfo {author} {\bibfnamefont {J.~R.}\ \bibnamefont
  {Ellis}}, \bibinfo {author} {\bibfnamefont {N.~E.}\ \bibnamefont
  {Mavromatos}}, \bibinfo {author} {\bibfnamefont {D.~V.}\ \bibnamefont
  {Nanopoulos}},\ and\ \bibinfo {author} {\bibfnamefont {S.}~\bibnamefont
  {Sarkar}},\ }\bibfield  {title} {\bibinfo {title} {{Tests of quantum gravity
  from observations of gamma-ray bursts}},\ }\href
  {https://doi.org/10.1038/31647} {\bibfield  {journal} {\bibinfo  {journal}
  {Nature}\ }\textbf {\bibinfo {volume} {393}},\ \bibinfo {pages} {763}
  (\bibinfo {year} {1998})},\ \Eprint {https://arxiv.org/abs/astro-ph/9712103}
  {arXiv:astro-ph/9712103} \BibitemShut {NoStop}%
\bibitem [{\citenamefont {Kifune}(1999)}]{Kifune:1999ex}%
  \BibitemOpen
  \bibfield  {author} {\bibinfo {author} {\bibfnamefont {T.}~\bibnamefont
  {Kifune}},\ }\bibfield  {title} {\bibinfo {title} {{Invariance violation
  extends the cosmic ray horizon?}},\ }\href {https://doi.org/10.1086/312057}
  {\bibfield  {journal} {\bibinfo  {journal} {Astrophys. J. Lett.}\ }\textbf
  {\bibinfo {volume} {518}},\ \bibinfo {pages} {L21} (\bibinfo {year}
  {1999})},\ \Eprint {https://arxiv.org/abs/astro-ph/9904164}
  {arXiv:astro-ph/9904164} \BibitemShut {NoStop}%
\bibitem [{\citenamefont {Peskin}\ and\ \citenamefont
  {Schroeder}(1995)}]{peskin1995introduction}%
  \BibitemOpen
  \bibfield  {author} {\bibinfo {author} {\bibfnamefont {M.}~\bibnamefont
  {Peskin}}\ and\ \bibinfo {author} {\bibfnamefont {D.}~\bibnamefont
  {Schroeder}},\ }\href {https://books.google.fr/books?id=EVeNNcslvX0C} {\emph
  {\bibinfo {title} {An Introduction To Quantum Field Theory}}},\ Frontiers in
  Physics\ (\bibinfo  {publisher} {Avalon Publishing},\ \bibinfo {year}
  {1995})\BibitemShut {NoStop}%
\bibitem [{\citenamefont {Trestini}\ and\ \citenamefont
  {Blanchet}(2023)}]{Trestini:2023wwg}%
  \BibitemOpen
  \bibfield  {author} {\bibinfo {author} {\bibfnamefont {D.}~\bibnamefont
  {Trestini}}\ and\ \bibinfo {author} {\bibfnamefont {L.}~\bibnamefont
  {Blanchet}},\ }\bibfield  {title} {\bibinfo {title} {{Gravitational-wave
  tails of memory}},\ }\href {https://doi.org/10.1103/PhysRevD.107.104048}
  {\bibfield  {journal} {\bibinfo  {journal} {Phys. Rev. D}\ }\textbf {\bibinfo
  {volume} {107}},\ \bibinfo {pages} {104048} (\bibinfo {year} {2023})},\
  \Eprint {https://arxiv.org/abs/2301.09395} {arXiv:2301.09395 [gr-qc]}
  \BibitemShut {NoStop}%
\bibitem [{\citenamefont {Pais}\ and\ \citenamefont
  {Uhlenbeck}(1950)}]{Pais:1950za}%
  \BibitemOpen
  \bibfield  {author} {\bibinfo {author} {\bibfnamefont {A.}~\bibnamefont
  {Pais}}\ and\ \bibinfo {author} {\bibfnamefont {G.~E.}\ \bibnamefont
  {Uhlenbeck}},\ }\bibfield  {title} {\bibinfo {title} {{On Field theories with
  nonlocalized action}},\ }\href {https://doi.org/10.1103/PhysRev.79.145}
  {\bibfield  {journal} {\bibinfo  {journal} {Phys. Rev.}\ }\textbf {\bibinfo
  {volume} {79}},\ \bibinfo {pages} {145} (\bibinfo {year} {1950})}\BibitemShut
  {NoStop}%
\bibitem [{\citenamefont {Bailey}\ \emph {et~al.}(2023)\citenamefont {Bailey},
  \citenamefont {Gard}, \citenamefont {Nilsson}, \citenamefont {Xu},\ and\
  \citenamefont {Shao}}]{Bailey2023}%
  \BibitemOpen
  \bibfield  {author} {\bibinfo {author} {\bibfnamefont {Q.~G.}\ \bibnamefont
  {Bailey}}, \bibinfo {author} {\bibfnamefont {A.~S.}\ \bibnamefont {Gard}},
  \bibinfo {author} {\bibfnamefont {N.~A.}\ \bibnamefont {Nilsson}}, \bibinfo
  {author} {\bibfnamefont {R.}~\bibnamefont {Xu}},\ and\ \bibinfo {author}
  {\bibfnamefont {L.}~\bibnamefont {Shao}},\ }\bibfield  {title} {\bibinfo
  {title} {{Classical radiation fields for scalar, electromagnetic, and
  gravitational waves with spacetime-symmetry breaking}},\ }\href@noop {} {\
  (\bibinfo {year} {2023})},\ \Eprint {https://arxiv.org/abs/2307.13374}
  {arXiv:2307.13374 [gr-qc]} \BibitemShut {NoStop}%
\bibitem [{\citenamefont {et~al.}(2023)}]{solpaper}%
  \BibitemOpen
  \bibfield  {author} {\bibinfo {author} {\bibfnamefont {N.~A.~N.}\
  \bibnamefont {et~al.}},\ }\bibfield  {title} {\bibinfo {title} {{In
  progress}},\ }\href@noop {} {\  (\bibinfo {year} {2023})}\BibitemShut
  {NoStop}%
\bibitem [{\citenamefont {Schwartz}(1978)}]{schwartz}%
  \BibitemOpen
  \bibfield  {author} {\bibinfo {author} {\bibfnamefont {L.}~\bibnamefont
  {Schwartz}},\ }\href@noop {} {\emph {\bibinfo {title} {Th\'{e}orie des
  distributions}}}\ (\bibinfo  {publisher} {Hermann},\ \bibinfo {year}
  {1978})\BibitemShut {NoStop}%
\bibitem [{\citenamefont {Blanchet}\ and\ \citenamefont
  {Faye}(2000)}]{Blanchet:2000nu}%
  \BibitemOpen
  \bibfield  {author} {\bibinfo {author} {\bibfnamefont {L.}~\bibnamefont
  {Blanchet}}\ and\ \bibinfo {author} {\bibfnamefont {G.}~\bibnamefont
  {Faye}},\ }\bibfield  {title} {\bibinfo {title} {{Hadamard regularization}},\
  }\href {https://doi.org/10.1063/1.1308506} {\bibfield  {journal} {\bibinfo
  {journal} {J. Math. Phys.}\ }\textbf {\bibinfo {volume} {41}},\ \bibinfo
  {pages} {7675} (\bibinfo {year} {2000})},\ \Eprint
  {https://arxiv.org/abs/gr-qc/0004008} {arXiv:gr-qc/0004008} \BibitemShut
  {NoStop}%
\bibitem [{\citenamefont {Marchand}\ \emph {et~al.}(2020)\citenamefont
  {Marchand}, \citenamefont {Henry}, \citenamefont {Larrouturou}, \citenamefont
  {Marsat}, \citenamefont {Faye},\ and\ \citenamefont
  {Blanchet}}]{Marchand:2020fpt}%
  \BibitemOpen
  \bibfield  {author} {\bibinfo {author} {\bibfnamefont {T.}~\bibnamefont
  {Marchand}}, \bibinfo {author} {\bibfnamefont {Q.}~\bibnamefont {Henry}},
  \bibinfo {author} {\bibfnamefont {F.}~\bibnamefont {Larrouturou}}, \bibinfo
  {author} {\bibfnamefont {S.}~\bibnamefont {Marsat}}, \bibinfo {author}
  {\bibfnamefont {G.}~\bibnamefont {Faye}},\ and\ \bibinfo {author}
  {\bibfnamefont {L.}~\bibnamefont {Blanchet}},\ }\bibfield  {title} {\bibinfo
  {title} {{The mass quadrupole moment of compact binary systems at the fourth
  post-Newtonian order}},\ }\href {https://doi.org/10.1088/1361-6382/ab9ce1}
  {\bibfield  {journal} {\bibinfo  {journal} {Class. Quant. Grav.}\ }\textbf
  {\bibinfo {volume} {37}},\ \bibinfo {pages} {215006} (\bibinfo {year}
  {2020})},\ \Eprint {https://arxiv.org/abs/2003.13672} {arXiv:2003.13672
  [gr-qc]} \BibitemShut {NoStop}%
\bibitem [{\citenamefont {Gel’fand}\ and\ \citenamefont
  {Shilov}(1964)}]{gelshi}%
  \BibitemOpen
  \bibfield  {author} {\bibinfo {author} {\bibfnamefont {I.~M.}\ \bibnamefont
  {Gel’fand}}\ and\ \bibinfo {author} {\bibfnamefont {G.~E.}\ \bibnamefont
  {Shilov}},\ }\href@noop {} {\emph {\bibinfo {title} {Generalized
  functions}}}\ (\bibinfo  {publisher} {Academic Press},\ \bibinfo {year}
  {1964})\BibitemShut {NoStop}%
\bibitem [{\citenamefont {Blanchet}(2014)}]{Blanchet:2013haa}%
  \BibitemOpen
  \bibfield  {author} {\bibinfo {author} {\bibfnamefont {L.}~\bibnamefont
  {Blanchet}},\ }\bibfield  {title} {\bibinfo {title} {{Gravitational Radiation
  from Post-Newtonian Sources and Inspiralling Compact Binaries}},\ }\href
  {https://doi.org/10.12942/lrr-2014-2} {\bibfield  {journal} {\bibinfo
  {journal} {Living Rev. Rel.}\ }\textbf {\bibinfo {volume} {17}},\ \bibinfo
  {pages} {2} (\bibinfo {year} {2014})},\ \Eprint
  {https://arxiv.org/abs/1310.1528} {arXiv:1310.1528 [gr-qc]} \BibitemShut
  {NoStop}%
\bibitem [{\citenamefont {Blanchet}\ \emph {et~al.}(2005)\citenamefont
  {Blanchet}, \citenamefont {Damour}, \citenamefont {Esposito-Farese},\ and\
  \citenamefont {Iyer}}]{Blanchet:2005tk}%
  \BibitemOpen
  \bibfield  {author} {\bibinfo {author} {\bibfnamefont {L.}~\bibnamefont
  {Blanchet}}, \bibinfo {author} {\bibfnamefont {T.}~\bibnamefont {Damour}},
  \bibinfo {author} {\bibfnamefont {G.}~\bibnamefont {Esposito-Farese}},\ and\
  \bibinfo {author} {\bibfnamefont {B.~R.}\ \bibnamefont {Iyer}},\ }\bibfield
  {title} {\bibinfo {title} {{Dimensional regularization of the third
  post-Newtonian gravitational wave generation from two point masses}},\ }\href
  {https://doi.org/10.1103/PhysRevD.71.124004} {\bibfield  {journal} {\bibinfo
  {journal} {Phys. Rev. D}\ }\textbf {\bibinfo {volume} {71}},\ \bibinfo
  {pages} {124004} (\bibinfo {year} {2005})},\ \Eprint
  {https://arxiv.org/abs/gr-qc/0503044} {arXiv:gr-qc/0503044} \BibitemShut
  {NoStop}%
\bibitem [{\citenamefont {Blanchet}\ \emph {et~al.}(2004)\citenamefont
  {Blanchet}, \citenamefont {Damour},\ and\ \citenamefont
  {Esposito-Farese}}]{Blanchet:2003gy}%
  \BibitemOpen
  \bibfield  {author} {\bibinfo {author} {\bibfnamefont {L.}~\bibnamefont
  {Blanchet}}, \bibinfo {author} {\bibfnamefont {T.}~\bibnamefont {Damour}},\
  and\ \bibinfo {author} {\bibfnamefont {G.}~\bibnamefont {Esposito-Farese}},\
  }\bibfield  {title} {\bibinfo {title} {{Dimensional regularization of the
  third postNewtonian dynamics of point particles in harmonic coordinates}},\
  }\href {https://doi.org/10.1103/PhysRevD.69.124007} {\bibfield  {journal}
  {\bibinfo  {journal} {Phys. Rev. D}\ }\textbf {\bibinfo {volume} {69}},\
  \bibinfo {pages} {124007} (\bibinfo {year} {2004})},\ \Eprint
  {https://arxiv.org/abs/gr-qc/0311052} {arXiv:gr-qc/0311052} \BibitemShut
  {NoStop}%
\bibitem [{\citenamefont {Bailey}\ \emph {et~al.}(2013)\citenamefont {Bailey},
  \citenamefont {Everett},\ and\ \citenamefont {Overduin}}]{Bailey:2013oda}%
  \BibitemOpen
  \bibfield  {author} {\bibinfo {author} {\bibfnamefont {Q.~G.}\ \bibnamefont
  {Bailey}}, \bibinfo {author} {\bibfnamefont {R.~D.}\ \bibnamefont
  {Everett}},\ and\ \bibinfo {author} {\bibfnamefont {J.~M.}\ \bibnamefont
  {Overduin}},\ }\bibfield  {title} {\bibinfo {title} {{Limits on violations of
  Lorentz Symmetry from Gravity Probe B}},\ }\href
  {https://doi.org/10.1103/PhysRevD.88.102001} {\bibfield  {journal} {\bibinfo
  {journal} {Phys. Rev. D}\ }\textbf {\bibinfo {volume} {88}},\ \bibinfo
  {pages} {102001} (\bibinfo {year} {2013})},\ \Eprint
  {https://arxiv.org/abs/1309.6399} {arXiv:1309.6399 [hep-ph]} \BibitemShut
  {NoStop}%
\end{thebibliography}%

\end{document}